\documentclass[preprint,preprintnumbers,amsmath,amssymb,prd,tightenlines
]{revtex4}
\usepackage{graphicx}
\begin{document}
\setlength{\voffset}{.5cm} 
\preprint{FAU-TP3-08-02}
\title{Canonical quantization of gauge fields in static space-times with applications to Rindler spaces}
\author{F. Lenz $^{{\rm a}}$}    \email{flenz@theorie3.physik.uni-erlangen.de} 
\author{K. Ohta $^{{\rm b}}$}    \email{ohta@nt1.c.u-tokyo.ac.jp}
\author{K. Yazaki $^{{\rm c}}$}  \email{yazaki@phys.s.u-tokyo.ac.jp}
\affiliation{$^{{\rm a}}$ Institute for Theoretical Physics III \\
University of Erlangen-N\"urnberg \\
Staudstrasse 7, 91058 Erlangen, Germany\\ \\
$^{{\rm b}}$ Institute of Physics \\ University of Tokyo\\ Komaba, Tokyo 153, Japan  \\ \\
$^{{\rm c}}$College of Arts and Sciences\\
Tokyo Woman's Christian University\\
Suginami-ku, Tokyo 167-8585, Japan\\
and\\ En'yo Radiation Laboratory, Nishina Accelerator Research Center, RIKEN,
Wako, Saitama 351-0198, Japan \\}
\pacs{11.15.-q, 04.62.+v, 11.10.K}
\date{August 20, 2008}
\begin{abstract}
The canonical quantization in the Weyl gauge of gauge fields in static space-times is presented. With an appropriate definition of transverse and longitudinal components of gauge fields, the Gau\ss\ law constraint is resolved explicitly for  scalar and spinor QED  and a complete non-perturbative solution is given for the  quantized Maxwell-field coupled to external currents. The formalism is applied in investigations of the electromagnetic field in Rindler spaces. The relation of creation and annihilation operators in Minkowski and Rindler spaces is established  and initial value problems associated with bremsstrahlung of a uniformly accelerated charge are studied. The peculiar scaling properties of scalar and  gauge theories in Rindler spaces are discussed and various quantities such as the photon condensate or the interaction energy of static charges and of scalar sources are computed.   
\end{abstract} 
\pacs{04.62.+v, 11.15.-q, 11.10.Kk }
\maketitle
\tableofcontents
\section{Introduction}
Over the past decade the study of quantum fields  in curved space-time has attracted considerable attention.  With the formulation of the AdS/CFT correspondence the possibility has emerged that theories such as QCD can  be formulated effectively as non-interacting  quantum field theories in curved space-time.   With  string theory inspired choices of the space-time metric  one has been able to describe non-perturbative phenomena such as confinement and chiral symmetry breaking with their associated phase transitions and a series of specific issues has been addressed concerning the spectrum  of QCD, the dynamics of Wilson loops or thermodynamic properties of the quark-gluon plasma such as its viscosity. Our studies are intended to improve and extend our understanding of the peculiar properties of the dynamics of quantized fields, in particular of gauge fields, in static space-times.

Studies of quantized gauge fields in curved  space-times  have been carried out almost exclusively in  the (general) covariant  Lorenz gauge.  The explicit covariance of the resulting formulation is achieved at the  expense of keeping redundant quantum fields. In static  space-times the existence of a timelike Killing vector singles out the time direction and  the Weyl (temporal) gauge becomes a natural choice for a canonical formulation of  gauge theories in terms of physical degrees of freedom only, e.\,g. in terms of transverse photons or gluons. In the first part of this work  we will carry out  the canonical quantization of gauge fields in arbitrary static space-times in the Weyl gauge with the Gau\ss\ law acting as a constraint on the space of physical states. The Gau\ss\ law  will be implemented explicitly for the Maxwell field coupled to external currents and  for scalar and  spinor QED. An essential element in this construction is the metric-dependent separation of vector fields into longitudinal and transverse components with the longitudinal degrees of freedom described by the eigenstates of the longitudinal conjugate momentum, i.\,e. of  the longitudinal electric field operator.  The coupling of the Maxwell field to external time dependent c-number currents is treated in detail and the time evolution operator is constructed without invoking perturbation theory. The coupling to charged scalar and spinor fields is also briefly discussed. First steps in the quantization of  non-Abelian gauge theories will be carried out  and the strategy for resolving the Gau\ss\ law constraint will be indicated for specific space-times.         

In the application of the general formalism we  will study the dynamics of  electromagnetic fields in Rindler space. Rindler spaces play a special role in the  study of quantum fields in curved space-times. On the one hand, a Rindler space can be viewed as  space-time under the influence of a static, homogeneous gravitational field  like the field close to the horizon of a Schwarzschild black hole. On the other hand, according to the equivalence principle,  a Rindler space can be be interpreted as part of a Minkowski space (Rindler wedge) as seen by a uniformly accelerated observer. This  equivalence offers the unique possibility to relate  explicitly the Heisenberg field operators   in Rindler space to those in  Minkowski space and to connect   creation and annihilation operators in the two spaces  via a Bogoliubov transformation. In this way  the Minkowski-space vacuum  is seen to differ from the Rindler-space vacuum \cite{FULL73,DAVI75,UNRU76} by  an acceleration induced thermal radiation of Rindler particles \cite{UNRU76}  (Unruh effect).

Besides important conceptional issues concerning e.g. the classical and quantum mechanical aspects of the Unruh radiation and its detection,  its relation to Hawking radiation, the Casimir effect in Rindler spaces  or the absence of radiation in the frame where a uniformly accelerated external charge is at rest (for a discussion of some of the open issues cf.\,\cite{FULL05}), investigations  have also addressed the phenomenological relevance and experimental verification of the Unruh effect (for a review cf.\,\cite{CRHM07}). Among various topics we mention the  depolarization of charged particles by acceleration (cf.\,\cite{BELE87}), the acceleration induced decay of particles \cite{MULL97,VAMA00} or the description of the thermalization  in relativistic heavy ion collisions in terms of  the Unruh effect \cite{KATU05}.   Experimental developments in atomic physics and optics afford  new opportunities for experimental investigations of the Unruh radiation.  Despite  the phenomenological relevance of electrodynamics most of the  investigations of the Unruh effect have been carried out in the context of scalar field theory. The rather few applications to the Maxwell field are mostly based on the quantization in the (covariant) Lorenz-gauge (cf.\,\cite{CADE77,TAGA86, HIMS921, AVSY02}). In related problems,  Gupta-Bleuler quantization in a modified Feynman gauge \cite{CRHM01} and the Weyl gauge in a semiclassical calculation \cite{COLE98} have been used.     
 
After a short review of the properties of quantized scalar fields in Rindler space,  we will apply  the general quantization scheme in the Weyl gauge  to electromagnetic fields in Rindler spaces and will construct explicitly the expansion of the gauge field in terms of the transverse eigenmodes of the Maxwell equation. We will identify a  symmetry related to  scale transformations which is peculiar to Rindler spaces and has important implications for the spectrum of scalar and gauge field theories.    The vector character of the gauge field and its gauge dependence make the relation between the Heisenberg field operators in inertial and accelerated frames intricate.  The central result of the analysis of this  relation  is  the expression of the  creation and annihilation operators of the transverse degrees of freedom   in Rindler space in terms of the corresponding operators in Minkowski space. Based on this result, a complete solution of initial value problems related to external currents will be given.  Equipped with these tools we address the issue of bremsstrahlung of charges at rest in the accelerated frame and  consider the case that the external charges are switched on adiabatically. The dynamics of the longitudinal degrees of freedom will be studied  in a calculation of the Coulomb interaction of two charges at rest in uniformly accelerated frames. In   comparison  to the dynamics of scalar fields we will find significant differences in the dynamics of both the longitudinal and  the  transverse components of the gauge fields. We will indicate  consequences of  our results for the structure of uniformly accelerated atoms and discuss  possible  implications for the phase structure of Yang-Mills theories and QCD in Rindler spaces. 
\section{Canonical  quantization of  the electromagnetic field  in  static space-times}
\subsection{The electromagnetic field  in Weyl gauge}
Here we carry out  the canonical quantization of the Maxwell field in a static but otherwise general $d+1$ dimensional space-time. A static  metric has the following structure ($i,j = 1,\ldots, d$)  
\begin{equation}
ds^2= g_{0 0}({\bf x})dx^0dx^0+ g_{i j}({\bf x})\,dx^{i} dx^{j} \,, \quad {\bf x}= (x^1,\ldots, x^d)\,.
\label{stame}
\end{equation}
The Lagrangian of the Maxwell field $A$ coupled to a conserved  external current $j$ is given by
\begin{equation}
{\cal L}   = \sqrt{|g|}\Big[- \frac{1}{4} g^{\mu \rho} g^{\nu \sigma}  (\partial_{\mu} A_{\nu} - \partial_{\nu} A_{\mu})
(\partial_{\rho} A_{\sigma} - \partial_{\sigma} A_{\rho}) -g^{\mu \nu} A_{\mu} j_{\nu}\Big]\,,
\label{mala}
\end{equation}
with the associated equations of motion
\begin{equation}
\partial_{\mu}\sqrt{|g|}g^{\mu \rho} g^{\lambda \sigma} (\partial_{\rho} A_{\sigma}- \partial_{\sigma}A_{\rho})   -\sqrt{|g|}\, j^{\lambda} = 0\, .
\label{coeq}
\end{equation}
The antisymmetry of the field-strength implies current conservation 
\begin{equation}
\frac{1}{\sqrt{|g|}}\,\partial_{\mu}\sqrt{|g|}\, j^{\mu} =0\,.
\label{cuco}  
\end{equation}
The canonical quantization is most conveniently carried out in the Weyl gauge
\begin{equation}
A_0(x^0,{\bf x})=0.
\label{Weyl}
\end{equation}
In this gauge, the Gau\ss\ law ($\lambda = 0$ in Eq.\,(\ref{coeq})) reads
\begin{eqnarray}
\label{gau}
\partial_{i} \big(\sqrt{|g|}\, g^{0 0}\, \partial_{0} A^{i}\big)
+\sqrt{|g|}\, j^{0} &=& 0 \, .
\label{emi}
\end{eqnarray}
The components of the conjugate momentum  $\Pi^{i}(x)$ and  the Hamiltonian  are given by (cf. Eq. (\ref{mala}))
\begin{equation}
\label{com}
\Pi^{i} = \frac{\partial {\cal L}}{\partial \partial_{0} A_{i}} = - \sqrt{|g|}\, g^{0 0}\, \partial_{0} A^{i} \, ,
\end{equation}
\begin{equation}
H = H_A+\int d^{\,d} x \sqrt{|g|} g_{i j}A^{i}j^{j} \,, \quad
%\end{equation*}
%\begin{equation}
H_A = \int d^{\,d} x  \Big\{-\frac{1}{2} \frac{g_{i j}}{\sqrt{|g|} \,g^{00}} \Pi^{i}\Pi^{j} + V[A] \Big\}  \, ,  \label{ham2}
\end{equation}
with the  energy density of the magnetic field 
\begin{eqnarray}
 V[A] &=& \frac{1}{4} \sqrt{|g|}  g^{ik} g^{jl}
(\partial_{i} A_{j} - \partial_{j} A_{i}) (\partial_{k} A_{l} - \partial_{l} A_{k}) \, .
 \label{magfi}
\end{eqnarray}
In the Weyl gauge, the Gau\ss\ law  (\ref{gau}) is not an equation of motion but  rather constrains the longitudinal electric field 
\begin{equation}
G= \partial_{i}\Pi^{i} - \sqrt{|g|} j^{0} =0\,.
\label{gaucc}
\end{equation}
The quantization in the Weyl gauge is standard. We postulate canonical commutation relations 
\begin{equation}
[\Pi^{i}(x^0, {\bf x}), A_{j}(x^0,{\bf x}^{\prime})] 
= \frac{1}{i}\delta_{i j } \delta({\bf x} - {\bf x}^{\prime}) \,.
\label{caco}
\end{equation}
No redundant degrees of freedom are present in this formulation though a residual gauge symmetry still exists. 
It is easy to verify that the Gau\ss\ law operator  (\ref{gaucc}) commutes with the Hamiltonian
\begin{equation}
  \label{gau1}
\left[G,H\right]=0\, .
\end{equation}
Thus,  the generator of time independent gauge transformations  $G$, can be diagonalized 
simultaneously with the Hamiltonian. One defines as the space of physical states $ |\Psi\rangle $ 
the space of eigenstates of $G$ with eigenvalue $0$,
\begin{equation}
  \label{gau2}
G|\Psi\rangle =0\, .
\end{equation}
\subsection{Transverse gauge fields}
The invariance under time independent gauge transformations  can be exploited  to further simplify the formalism. The strategy now depends on the details of the theory  as well as on the properties of the metric. In the canonical formulation of the Maxwell theory or of QED in flat space the reduction  to ``Coulomb gauge'' is  natural. In this formulation static charges are not coupled to the radiation field. We will  show in the following sections, that this implementation of the residual gauge symmetry can be achieved  for the Maxwell field coupled to an external current and for QED in an arbitrary static space-time by an appropriate generalization of the  concept of longitudinal and transverse fields. To this end we  introduce a decomposition of vector fields into  longitudinal and transverse components which generalizes the flat space definition to static curved  space-time and define   
 the following projection operators on transverse and longitudinal fields 
\begin{equation}
  \label{proj1}
P^{i}_{\;j} = g^{i}_{\;j} + \partial^{i} \; \frac{1}{\Delta} \;
\partial_{j} , \quad  Q^{i}_{\;j} = - \partial^{i} \; \frac{1}{\Delta}\partial_{j}\, ,
\end{equation}
with the Laplacian 
\begin{equation}
  \label{lapla}
\Delta = - \partial_{i} \partial^{i}\, .
\end{equation}
The projection operators satisfy the standard requirements 
\begin{equation}
  \label{proj2}
P^{i}_{\;j}\,P^{j}_{\;k} = P^{i}_{\,k}\, , \quad Q^{i}_{\;j}\,Q^{j}_{\;k} = Q^{i}_{\,k}\,, \quad  P^{i}_{\,k} + Q^{i}_{\,k} = \delta_{ik}\, ,\quad P^{i}_{\;j}\,Q^{j}_{\;k} = 0\, ,
\end{equation}
and 
\begin{equation}
  \label{proj3}
\partial_{i}P^{i}_{\;j}= 0,\quad \partial_{i}Q^{i}_{\;j} = \partial_{j}\, .
\end{equation}
We also note  the hermiticity of the Laplacian (\ref{lapla})
$$ \int d^d x f({\bf x})\, \Delta g({\bf x}) =  \int d^d x  \Delta f({\bf x}) \,g({\bf x}) \,.$$
As in flat space, these properties guarantee the orthogonality of longitudinal and transverse components of vector fields 
\begin{equation}
  \label{orthg}
\int d^d x P^i_{\,j} V^{j}({\bf x})\,  Q_{i\,k} W^{k}({\bf x}) = 0 \,,   
\end{equation}
and 
\begin{equation}
  \label{intpro}
\int d^d x P^i_{\,j} V^{j}({\bf x})\,  W_{i}({\bf x}) = \int d^d x P^i_{\,j} V^{j}({\bf x}) \,  P_{i\,k} W^{k}({\bf x})\,.    
\end{equation} 
We denote the transverse components of gauge and electric fields by
\begin{equation}
 \hat{A}^{i}= P^{i}_{\; j}\,A^{j}\, ,\quad \hat{\Pi}^i= P^{i}_{\;j}\,\Pi^{j}\,.
\label{traug}
\end{equation}
This definition of the transverse fields  and the canonical commutation relations 
%(\ref{caco}) 
imply
\begin{eqnarray}
%& & 
[\hat{\Pi}^{i}(x^0,{\bf x}), \hat{A}_{j}(x^0,{\bf x}^{\prime})] = \frac{1}{i}\Big(g^i_{\,\, j}\delta({\bf x}-{\bf x}^{\prime}) - \partial^i \partial^{\prime}_j \, D( {\bf x},{\bf x}^{\prime})\Big)\,,
\label{patr}
\end{eqnarray}
with 
\begin{equation}
D({\bf x},{\bf x}^{\prime}) = \langle {\bf x} |\frac{1}{\Delta}| {\bf x}^{\prime}\rangle\,. 
\label{dxxp}
\end{equation}
Evidently, this is the generalization of the transverse commutation relations of electric and gauge fields in flat space.
\subsection{Maxwell field and external currents}
\subsubsection{Decoupling of external currents} \label{sec:decpl}
We first consider the Maxwell field  coupled to an external current. This system is described by the Hamiltonian (\ref{ham2}) 
\begin{equation}
H =\int d^{\,d} x\Big\{-\frac{1}{2} \frac{g_{i j}}{\sqrt{|g|} \,g^{00}} \Pi^{i}\Pi^{j} +  V[\hat{A}]+\sqrt{|g|} g_{i j}A^{i}j^{j}\Big\} \,.
\label{HA2}
\end{equation} 
In this case of a c-number current,  the terms linear in the gauge and electric field operators can be eliminated by a unitary transformation and the Hamiltonian can be formulated in terms of  transverse degrees of freedom only. 
We define a transformation which shifts fields and conjugate momenta
\begin{equation}
W = e^{i\Phi(t)}=  e^{i \int d^{\,d} x (- \Pi^{k} \alpha_{k} + A^{k} p_{k})} \,, \quad
W^{\dagger} A_{i} W = A_{i} + \alpha_{i} \,, \quad W^{\dagger} \Pi^{i} W = \Pi^{i} + p^{i}\,.
\label{shif}
\end{equation} 
The Hamiltonian transforms  as follows
\begin{equation}
  \label{HWH}
H^{\prime}=W^{\dagger} H W - i W^{\dagger} \partial_{t} W \,,  
\end{equation}
and the Gau\ss\ law in the transformed space of physical states becomes 
\begin{equation}
0= W^{\dagger}\Big( \partial_{i} \Pi^{i} - \sqrt{|g|}\; j^{0}\Big) W| \psi^{\prime} \rangle =\big( \partial_{i} \Pi^{i} - \sqrt{|g|} \; j^{0} + \partial_{i} p^{i} \big )| \psi^{\prime} \rangle\,.
\label{trgl}
\end{equation}
Acting on the  transformed states 
\begin{equation}
 | \psi^{\prime} \rangle = W^{\dagger} | \psi \rangle\,,
\label{trast}
\end{equation}
the electric field operator is  decomposed accordingly into transverse and longitudinal components 
\begin{equation}
\Pi^{i} | \psi^{\prime} \rangle
= \Big(\hat{\Pi}^{i} - \partial^{i} \; \frac{1}{\Delta} \; \big ( \sqrt{|g|} \; j^{0} - 
\partial_{k} p^{k}\big) \Big ) | \psi^{\prime} \rangle\,.
\label{trgl2}
\end{equation}
For computing the time derivative we rewrite $W$
\begin{equation}
W = e^{- i \int d^{\,d} x \Pi^{k} \alpha_{k}} \; e^{i \int d^{\,d}x A_{k} p^{k}} \; 
e^{ \frac{i}{2} \int d^{\,d}x \alpha_{k} p^{k}}\,.
\label{utw}
\end{equation}
A straightforward calculation yields the  following decomposition of the  Hamiltonian 
\begin{equation}
H^{\prime} = H_0 + H_{1}+ h(t)\,.
\label{hutp}
\end{equation}
into  the Hamiltonian of the uncoupled Maxwell field  $ H_0$ 
\begin{equation}
H_0= \int d^{\,d}x \Big ( - \frac{1}{2} \, g_{ij} \; \frac{1}{\sqrt{|g|}\,g^{00}}\; 
\hat{\Pi}^{i} \hat{\Pi}^{j} + V [\hat{A}] \Big)\,,
\label{hp2}
\end{equation}
the contribution  $H_1$,  linear in  $A$ and $\Pi$ 
\begin{equation}
H_1=\int d^{\,d}x \, \Big [ \sqrt{|g|}\, A_{k} j^{k} + \frac{1}{2} \, \sqrt{|g|} \, g^{ik} g^{j \ell} f_{ij} F_{k \ell} + \dot{p}^{i} A_{i}- \hat{\Pi}^{i} \dot{\alpha}_{i}-\frac{1}{\sqrt{|g|}\,g^{00}} \, \hat{\Pi}^{i} 
\big (\hat{p}_{i} - \partial_{i} \frac{1}{\Delta} \, \sqrt{|g|}\, j^{0} \big ) \Big ]\,, \\
\label{hp1}
\end{equation}
and the c-number contribution to the Hamiltonian $h(t)$
\begin{eqnarray}
\hspace*{-0.8cm} h(t) & = & \int d^{\,d}x \Big\{\sqrt{|g|} \alpha_{i} j^{i} + \frac{1}{4} \sqrt{|g|} \, g^{ik} g^{j \ell} 
f_{ij} f_{k \ell}- \frac{1}{2 \sqrt{|g|}\,g^{00}} \Big ( \hat{p}^{i} - \partial^{i} \frac{1}{\Delta} \; \sqrt{|g|}\, j_{0}
\Big )  \nonumber\\
 &&\cdot\Big ( \hat{p}_{i} - \partial_{i} \frac{1}{\Delta} \, \sqrt{|g|} \, j_{0} \Big ) +\dot{\alpha}_i\partial^i\frac{1}{\Delta}(\sqrt{|g|}j^0-\partial_k p^k)  %
+ \frac{1}{2} \, \big (\dot{p}^{i} \alpha_{i} - p^{i} \dot{\alpha}_{i} \big )\Big\} \,.  \label{ktan0}
\end{eqnarray}
We have introduced  the c-number field strength of the auxiliary gauge fields $\alpha$ 
\begin{displaymath}
f_{ij} = \partial_{i} \alpha_{j} - \partial_{j} \alpha_{i}\, ,
\end{displaymath} 
and have denoted the time derivatives of the auxiliary gauge  ($\alpha_i$)  and   electric ($p^i$) fields by a dot. The auxiliary fields    are  used  to make $H_1$ (\ref{hp1})  vanish. The terms linear in  $A_i$ do not contribute if
\begin{equation}
\sqrt{|g|} \, j^{i} + \partial_{l} \big(\sqrt{|g|}\, g^{ik} g^{jl} f_{kj}\big) + \dot{p}^{i} = 0 \,,        \label{GLC0}
\end{equation}
or written in terms of transverse and longitudinal components 
\begin{equation}
\label{GLC}
p^i_{(\ell)} = -\partial^i\frac{1}{\Delta} \sqrt{|g|} j^0- E^i_{(\ell)} ({\bf x})\,,\quad \hat{\dot{p}}^i= -\sqrt{|g|} \, j^{i} - \partial_{l} \big(\sqrt{|g|}\, g^{ik} g^{jl} f_{kj}\big) +\partial^i\frac{1}{\Delta}\sqrt{|g|} \partial_0 j^0\,\,.
\end{equation}
The longitudinal electric field if acting on the transformed states (cf.\,Eq.\,(\ref{trgl2})) is given  by the integration constant $E^i_{(\ell)}$  
\begin{equation}
\Pi^{i} | \psi^{\prime} \rangle
= \Big(\hat{\Pi}^{i}+E^i_{(\ell)}({\bf x}) \Big ) | \psi^{\prime} \rangle\,.
  \label{inco}
\end{equation}
Vanishing of the terms linear in $\hat{\Pi} $ requires 
\begin{equation}
\dot{\alpha}_{i} = \partial_i \
\dot{\Omega} - \frac{1}{\sqrt{|g|}\, g^{00}} \; \Big ( \hat{p}_{i} - \partial_{i} \frac{1}{\Delta} \, \sqrt{|g|} 
\, j^{0} \Big )\,,      \label{eqch0}
\end{equation}
where $\Omega$  is an arbitrary function. Here the gauge freedom arises, since the projection on the transverse conjugate momentum $\hat{\Pi}$ leaves the longitudinal component of $\alpha$ undetermined.  
We differentiate this equation with respect to time,  use Eq.\,(\ref{GLC0}) and current conservation (Eq.\,(\ref{cuco})) to derive the following wave equation 
\begin{equation}
0 = - \sqrt{|g|} \, g^{00} \ddot{\alpha}^{i}  + 
\sqrt{|g|} \, g^{00} \partial^i\ddot{\Omega}+ \sqrt{|g|} \, j^{i} + \partial_{l} \big(\sqrt{|g|}\, g^{ik} g^{jl} f_{kj}\big) \,. \label{eqc0}
\end{equation}
With the gauge choice $\Omega=0$,  this wave equation is nothing else than the  equation of motion (\ref{coeq}) in the Weyl gauge.\\
Thus provided the auxiliary functions are chosen to satisfy the equations of motion (\ref{GLC0}) and (\ref{eqch0}) the Hamiltonian is a sum of the Hamiltonian of the Maxwell field formulated in terms of transverse degrees of freedom and of a c-number contribution accounting  for the effect of external currents
\begin{equation}
  \label{h0}
  H'= H_0 + h(t)\,.
\end{equation}
For  solutions of the equation of motion (\ref{GLC0}) and (\ref{eqch0})  the c-number part  $h(t) $ (\ref{ktan0}) simplifies 
 \begin{equation}
 \label{hx0}
 h(t) = \frac{1}{2} \int d^{d} x \Big\{\sqrt{|g|} \alpha_{k}j^k - \dot{\alpha}_iE^i_{(\ell)}+\dot{\Omega} \sqrt{|g|} j^0\Big\} \,.  
 \end{equation}
It is advantageous to eliminate the longitudinal component of the auxiliary field with the following choice of the gauge function $\Omega$ 
\begin{equation}
 \label{omtr}\dot{\Omega} = -\frac{1}{\Delta}\partial_i\Big( \frac{1}{\sqrt{|g|} g^{00}} ( \hat{p}^i-\partial^i\frac{1}{\Delta}\sqrt{|g|} j^0)\Big)\,,\quad h(t) = \frac{1}{2} \int d^{d} x \Big\{\sqrt{|g|} \alpha_{k}j^k  +\dot{\Omega} \sqrt{|g|} j^0\Big\} \,. 
\end{equation}
Besides the standard coupling of the transverse gauge field to the current and the   instantaneous Coulomb interaction energy $h_C(t)$,  an additional term arises from the coupling of the charges to the transverse electric field
\begin{eqnarray}
 \label{omtr2} h(t) &=& -\frac{1}{2} \int d^{d} x \Big\{\sqrt{|g|} \hat{\alpha}_{k}j^k + \sqrt{|g|} j^0\frac{1}{\Delta}\partial_i\frac{1}{\sqrt{|g|} g^{00}}  \hat{p}^i\Big\} + h_C(t) \,,\\
h_C(t) &=& \hspace{.3cm}\frac{1}{2} \int d^{d} x  \sqrt{|g|} j^0\frac{1}{\Delta}\partial_i\frac{1}{\sqrt{|g|} g^{00}}\partial^i\frac{1}{\Delta}\sqrt{|g|} j^0 \,.
\label{omtr3} 
 \end{eqnarray}
The appearance of  the  equations of motion (\ref{eqc0})  of the Maxwell field coupled to an external current  for the auxiliary fields had to be expected. In the classical context our procedure amounts to decompose the field in a homogeneous solution ($A_i\,,\Pi^i$) of the Maxwell equations and an inhomogeneous solution ($\alpha_i\,,p^i$).

The  Maxwell equations for the auxiliary fields  $\alpha_i$ and $p^i$ imply  a  useful identity for the exponent $\Phi(t)$ of the  unitary transformation (\ref{shif}). We decompose  the exponent of $W$  into transverse and longitudinal contributions 
\begin{equation}
\Phi (t)  =   \int  d^{\,d}x \big ( - \Pi ^{k} \alpha_{k} + p^{k}  A_{k}\big ) =  \hat{\Phi}(t) + \Phi_{(\ell)}(t) \,,   
\end{equation}
use  Eq.\,(\ref{GLC}) to rewrite $ \Phi_{(\ell)}(t) $, transform to Heisenberg operators  and derive with the help of the  equations of motion  (\ref{coeq}) and (\ref{eqc0}) 
\begin{equation}
\frac{d}{dt}\hat{\Phi}_{H} 
 = - \int d^{\,d} x \hat{A}_{H i} \sqrt{|g|} j^i+\int d^{\,d}x\hat{\Pi}_H^i \frac{1}{\Delta} \sqrt{|g|} j^0\, \partial_i \frac{1}{\sqrt{|g|} g^{00}} \,\,. 
\label{dpht}
\end{equation}
In the first term, only the  transverse component of the current  
\begin{equation}
  \label{trcu}
\hat{J}^{i}= (g^i_{k}+\partial^i\frac{1}{\Delta}\partial_k)\sqrt{|g|}\, j^{k} \,
\end{equation}
couples to the transverse component of the gauge field (cf.\,Eq.\,(\ref{intpro})). 
The second term couples static charges to the radiation field. It arises since in general,  in a static metric, like in the Schwarzschild  or the standard AdS metric 
\begin{equation}
ds^2=\frac{1}{u^2}\big(dx^{0\,\,2} -\sum_{i=1}^{d-1} dx^{i\,\,2}-du^2 \big)
\label{adsm} 
\end{equation}
the ratio   of  conjugate momentum to electric field varies in space.   
\\
The above expressions simplify if the coordinates can be chosen such that the metric  satisfies  
\begin{equation}
\partial_i (\sqrt{|g|} g^{00}) = 0 \,.
\label{rili} 
\end{equation}
In this case, as  in flat space-time,  the equation of motion (\ref{eqc0}) 
\begin{equation}
0 = - \sqrt{|g|} \, g^{00}\hat{\ddot{\alpha}}^{i}  + \hat{J}^i + \partial_{l} \big(\sqrt{|g|}\, g^{ik} g^{jl} f_{kj}\big) \,, \label{eqri}
\end{equation}
does not couple radiation field and static charges (cf.\,Eq.\,\ref{dpht})). The expression for the energy (cf.\,Eq.\,(\ref{omtr2})) contains the standard coupling of transverse gauge field and current and the electrostatic interaction (cf.\,Eq.\,(\ref{omtr3}))    
\begin{equation}
 \label{hri} h(t) = -\frac{1}{2} \int d^{d} x \sqrt{|g|} \hat{\alpha}_{k}j^k + h_C(t)\,,
 \end{equation}
\begin{equation}
  \label{coul}
 h_C(t)= -\frac{1}{2}\int d^dx \sqrt{|g({\bf x})|}\int d^dx^{\prime} \sqrt{|g({\bf x}^{\prime})|} j^0(t,{\bf x}) D({\bf x}, {\bf x}^{\prime}) j^0(t,{\bf x}^{\prime}) \,.  
\end{equation}
The transverse part $\hat{\Phi}$ of the exponent of the unitary transformation is given by the coupling of the transverse gauge field to the current 
\begin{equation}
\frac{d}{dt}\hat{\Phi}_{H} 
 = - \int d^{\,d} x \hat{A}_{H i} \hat{J}^{i}\,.
\label{dpht2}
\end{equation}
Presence or absence of coupling of the time component of the current to the radiation field is not an intrinsic property of the particular static space-time. It depends on the choice of the coordinates.  
If  the condition (\ref{rili}) is not fulfilled one may construct a time-independent coordinate transformation $x^i \to x^{\prime\,i}$ which decouples charges from the radiation field,  provided it  satisfies  the constraint
\begin{equation}\Big|\,\text{det}\,\frac{\partial\big(x^{\prime\, 1}\!\!\!,\ldots, x^{\prime\,d}\big)}{\partial\big(x^{1},\ldots, x^{d}\big)}\Big|= g^{00}\sqrt{|g|}\,,
\label{dxxp2} 
\end{equation}
up to a multiplicative constant. 
For instance, as is easily verified,  the coordinate transformation 
$$(t,r,\vartheta,\varphi) \to (t,\rho,\sigma,\varphi)\,\quad \text{with}\quad 
\rho=\int^rd r^{\prime} \frac{r^{\prime\,3}}{r^{\prime}-2M}\,,\quad \sigma= \cos(\vartheta)$$
applied to the Schwarzschild-metric \cite{MITW70} with Schwarzschild-radius $2M$ or the coordinate transformation
$$(x^{\mu},u) \to (x^{\mu},v=u^{-d+2})$$
applied to the AdS metric (\ref{adsm})  
decouple static charges and radiation field.  This decoupling does not necessarily simplify the calculation. In the case of the Schwarzschild metric, for instance, the  coordinate transformation  cannot be inverted in closed form which prevents an explicit formulation of the dynamics.  Another simple example for such a coordinate transformation will be discussed for the Rindler space in Sec.\,\ref{cafo}.      
\\
For the sake of comparison we compile here the relevant results for a scalar field coupled to an external source 
\begin{equation}
{\cal L}= \sqrt{|g|}\Big\{\frac{1}{2}\big[ g^{\mu\nu}\partial_{\mu}\phi\,\partial_{\nu}\phi -m^2\phi^2 \big]-\rho\phi\Big\}.
\label{lasc}
\end{equation}
In analogy with Eq.\,(\ref{shif}) we define the following unitary transformation \begin{equation}
  \label{unsc}
  w=e^{i\varphi(t)}\,,\quad \varphi(t) = \int d^{\,d}x (-\pi\alpha+\phi p) =  \int d^{\,d}x \sqrt{|g|} g^{00} (-\dot{\phi}\alpha+\phi\dot{\alpha}), 
\end{equation}
with the conjugate pairs of field operators $\phi,\, \pi$ and auxiliary fields $\alpha, p$ solving the inhomogeneous field equations
\begin{equation}
  \label{scem}
  \sqrt{|g|}g^{00} \ddot{\alpha} +\partial_i (\sqrt{|g|} g^{ij} \partial_j\alpha)+\sqrt{|g|} m^2 \alpha +\sqrt{|g|}\rho = 0 \,. 
\end{equation}
The exponent $\varphi(t)$  satisfies
\begin{equation}
  \label{unsc2}
 \frac{d}{dt} \varphi(t) = -\int d^{\,d}x \sqrt{|g|} \rho \phi\,.
\end{equation} 
After decoupling the external source by the  transformation $w$, the Hamiltonian (cf.\,Eq.(\ref{HWH})) is given by
\begin{equation}
  \label{hxs1}
  H^{sc\,\prime} = H_0^{sc} + h^{sc}(t)\,,
\end{equation} 
with 
\begin{equation}
  \label{hxs2}
  h^{sc}(t) = \frac{1}{2} \int d^{\, d} x \sqrt{|g|}\rho\, \alpha\,.
\end{equation} 
\subsubsection{Time evolution}\label{sec:tiev}
To illustrate in this general context the consequences of our results we consider a typical initial value problem as it arises when external sources are present. More detailed issues will be addressed later in the application to Rindler space.  

We assume that the external current is vanishing beyond some time
in the past
\begin{displaymath}
j^{\mu} (t, {\bf x}) = 0 \quad {\rm for} \quad t \le t_{0}\,.
\end{displaymath}
We calculate the time evolution of the state $|\psi(t_0)\rangle$ by transforming this state at $t_0$ into the primed basis (Eq.\,(\ref{trast})), let it evolve under the influence of $H^\prime$\,(\ref{h0}) and transform back to the unprimed basis at time $t$. We furthermore require  the auxiliary fields to satisfy the initial conditions 
$$\alpha^i(t_0,{\bf x}) = p^i(t_0,{\bf x}) =0\,.$$
 Thus the time evolution is given by
\begin{equation}
|\psi (t) \rangle = W(t) \, e^{- i H_{0} (t-t_{0})-i\int^t_{t_0} dt^{\prime} h(t^{\prime})} \, | \psi (t_{0}) \rangle \, ,
\label{tev}
\end{equation}
and the expectation value of a gauge invariant observable ${\cal O} (\Pi^{i}, \hat{A}_{i}) $  by 
\begin{displaymath}
\langle \psi (t)| {\cal O} (\Pi, \hat{A}) | \psi (t) \rangle = \langle \psi (t_{0}) | {\cal O}_{H} 
 | \psi (t_{0}) \rangle \,,
\end{displaymath}
with the Heisenberg operator 
\begin{eqnarray}
{\cal O}_{H} & = &  e^{i H_{0} (t-t_{0})} W^{\dagger} (t) {\cal O} (\Pi, \hat{A}) W (t) \, 
e^{- i H_{0} (t-t_{0})}  \nonumber \\
& = & {\cal O} \Big (\hat{\Pi}^{i}_{H} +\hat{p}^{i} - \partial^{i} \, \frac{1}{\Delta} \; \sqrt{|g|} \; j^{0}, 
\; \hat{A}_{iH} + \hat{\alpha}_{i} \Big )\,,
\label{heis}
\end{eqnarray}
where we have used  the transformation properties (\ref{shif}) of gauge and electric fields,  the Gau\ss\ law in the original basis (\ref{gaucc})  and, in integrating the equation of motion (\ref{GLC}) of the longitudinal electric field,  we have dropped a time independent integration constant.
The Heisenberg operators are defined with respect to $H_{0} $
\begin{displaymath}
\begin{array}{l}   \hat{\Pi}^{i}_{H}   \\ \hat{A}_{i,H}  \end{array}
= e^{i H_{0} (t-t_{0})} \; 
\begin{array}{l}  \hat{\Pi}^{i}  \\  \hat{A}_{i}   \end{array}  \; 
e^{- i H_{0} (t-t_{0})}  \,.
\end{displaymath}
Expression (\ref{heis}) summarizes the results of our discussion. Gauge and electric fields are given by the sum of the standard operator valued solution of the 
homogeneous Maxwell-equations and a special classical solution of the inhomogeneous Maxwell equation. 
\subsection{Quantum Electrodynamics}
\subsubsection{Scalar QED}
Here we briefly discuss the necessary modifications of the formalism if the gauge field is coupled to a charged matter field rather than to an external current. We first consider the case of a  scalar charged matter field with the action of the coupled  system  given by
\begin{equation}
S= \int d^{\,d+1}x \sqrt{|g|} \Big [ - \frac{1}{4}\; (\partial^{\mu} A^{\nu} - \partial^{\nu} A^{\mu}) (\partial_{\mu}
A_{\nu} - \partial_{\nu} A_{\mu}) + ( \partial^{\mu} + i e A^{\mu}) \phi^{\dagger} 
(\partial_{\mu} - i e A_{\mu}) \phi - m^{2} \phi^{\dagger} \phi  \Big ]\,.\\
\label{qedsc}
\end{equation}
The quantization of the gauge field in the Weyl gauge proceeds as before. The canonical momentum of the scalar field is given by 
\begin{displaymath}
\pi = \frac{\partial {\cal L}}{\partial (\partial_{0} \phi)} = \sqrt{|g|}\, g^{00}\,  \partial_{0} \phi^{\dagger} \,,
\end{displaymath} 
and the Gau\ss\ law constraint keeps its form (\ref{gaucc})
\begin{equation}
G|\psi_p\rangle = \Big(\partial_{i}\Pi^{i} - \sqrt{|g|}\, j^{0}\,)|\psi_p\rangle =0\,,
\label{gaucs}
\end{equation}
with the following definition of the charge density
\begin{equation} j^{0} = -i\frac{e}{\sqrt{|g|}} \; 
 (\phi^{\dagger} \pi^{\dagger} - \pi \phi) \,. 
\label{j0sc}
\end{equation}
The resulting Hamiltonian  in the Weyl gauge (cf. Eq.\,(\ref{ham2}))
\begin{eqnarray*}
H  = H_A+\int d^{\,d} x \Bigg\{ \frac{1}{\sqrt{|g|}\, g^{00}} \; \pi^{\dagger} \pi - \sqrt{|g|} \;\Big[ (\partial^{i} 
+ i e A^{i}) \phi^{\dagger} (\partial_{i} - i e A_{i}) \phi - m^{2} \phi^{\dagger} \phi \Big]\Bigg\}
\end{eqnarray*}
still contains  longitudinal gauge field components. As above they can be eliminated by an  unitary gauge fixing transformation which we choose as follows 
\begin{equation}
w = \exp\Big\{{i\int d^{d}  x \sqrt{|g|} j^{0} \frac{1}{\Delta} \partial_{i} A^{i}}\Big\}\,, 
\label{ugft0}
\end{equation}
with $j^0$ given in (\ref{j0sc}).  
This transformation acts on both the gauge and matter field. 
 The  transformed electric field reads  
\begin{equation}
\Pi^{\prime \,i}=w^{\dagger} \Pi^{i} w = \Pi^{i} - \partial^i \frac{1}{\Delta} \sqrt{|g|} \, j ^{0}\,, 
\label{UG}
\end{equation}
and thus in the space of the transformed physical states
$$ |\psi_{p}^{\prime} \rangle = w |\psi_p\rangle \,,$$
the Gau\ss\ law operator vanishes
\begin{equation}
 G ^{\prime}| \psi_{p}^{\prime} \rangle = \partial_i \Pi^i| \psi_{p}^{\prime}\, \rangle = 0  \,.
\label{gp}
\end{equation}
The transformation $w$ rotates the phase of the matter field, with the gauge function containing the longitudinal field operator
\begin{displaymath}
w^{\dagger} \phi\, w = e^{- i e \frac{1}{\Delta} \partial_{\beta} A^{\beta} } \phi \,, \quad 
w^{\dagger} \pi w = e^{i e \frac{1}{\Delta} \partial_{\beta} A^{\beta}}\pi\,,
\end{displaymath}
and eliminates the longitudinal component of the gauge field from the covariant derivative 
\begin{eqnarray*}
w^{\dagger} \big(\partial_{\alpha} - i e A_{\alpha}\big) \phi\, w  =   e^{- i e \frac{1}{\Delta} \partial_{\beta} A^{\beta}} \big(\partial_{\alpha} - i e \hat{A}_{\alpha}\big) \phi\,.
\end{eqnarray*} 
Unlike in the case of the Maxwell field coupled to an external current, here the unitary transformation is time independent and we find for the transformed Hamiltonian
\begin{eqnarray}
&&H^{\prime}= w^{\dagger} H w  
=H_A^{\prime} + h_C \nonumber \\
&&+ \int d^{\,d} x\Big\{\frac{1}{\sqrt{|g|}\, g^{00}} \, \pi^{\dagger} \pi - \sqrt{|g|}\Big[ (\partial^{i} 
+ i e \hat{A}^{i}) \phi^{\dagger} (\partial_{i} - i e \hat{A}_{i}) \phi - m^{2} \phi^{\dagger} \phi \Big]\Big\}
\,,
\label{haprsc}
\end{eqnarray}
with the transformed Hamiltonian of the Maxwell field (Eq.\,(\ref{ham2}))
\begin{equation}
H_A^{\prime}= w^{\dagger}H_Aw=\int d^{\,d} x\Big\{-\frac{1}{2} \frac{g_{i j}}{\sqrt{|g|} \,g^{00}} \hat{\Pi}^{i}\hat{\Pi}^{j} +j^0 \sqrt{|g|} g_{ij}\big(\partial^i\frac{1}{\sqrt{|g|} g^{0 0}}\big)\frac{1}{\Delta}\hat{\Pi}^{j}
%\nonumber \\ &&
+ V[\hat{A}] \Big\}
\,,
\label{hasc}
\end{equation}
and the Coulomb-energy operator
\begin{equation}
\label{fesc2}
h_{C}=-\frac{1}{2}  \int d^{\,d} x \frac{g^{ij}}{\sqrt{|g|} \,g^{0 0}} \big(\partial_i \frac{1}{\Delta}\sqrt{|g|} j^0)\big(\partial_j \frac{1}{\Delta}\sqrt{|g|} j^0) \,.
\end{equation}
We thus have achieved a canonical formulation of scalar QED  free of redundant degrees  of freedom. We note that also in this dynamical treatment of the current,  static charges do couple to the transverse electric field  in general static space-times.
\subsubsection{Spinor QED} 
For the coupled system of the Maxwell and a charged spin $1/2$ field with  action  
\begin{equation}
S = \int d^{\,d}x \sqrt{|g|} \Big \{ - \frac{1}{4}\; (\partial^{\mu} A^{\nu} - \partial^{\nu} A^{\mu}) (\partial_{\mu}
A_{\nu} - \partial_{\nu} A_{\mu}) + \bar{\psi}\big[\gamma^{\mu}i(\nabla_{\mu} -i e A_{\mu})-m \big]\psi  \Big\}\, ,
\label{qedsp}
\end{equation}
we proceed analogously. The space-time dependent $\gamma$ matrices satisfy the anticommutation relations
\begin{equation}
\{\gamma^{\mu}, \gamma ^{\nu}\}= 2 g^{\mu\nu} \,,
\label{codir}
\end{equation}
and the covariant derivative $\nabla_{\mu}$ 
is given by
$$\nabla_{\mu} = \partial_{\mu} -\Gamma_{\mu}$$
with $\Gamma_{\mu}$ determined by the ``spin connection'' \cite{BRWH57}. Here we neither need the explicit  expressions for the $\gamma$ matrices nor for $\Gamma_{\mu}$ (for details see  for instance \cite{NAKA90}) and  only mention that $\Gamma_{\mu}$ can be chosen time-independent for static metrics. 
From the action (\ref{qedsp}) we read off the expression for the time component of the current (cf. Eq.\,(\ref{mala})) 
\begin{equation} j^{0} = -e g^{00}\,\psi^{\dagger}\, \psi\,. 
\label{j0sp}
\end{equation}
The conjugate momentum is given by
$$ \pi = i \sqrt{|g|} g^{00} \psi^{\dagger}\,, $$
and the following equal-time anticommutation relations are postulated
\begin{equation}
\{\psi_{\alpha}(x^0, {\bf x}), \psi_{\beta}(x^0,{\bf x}^{\prime})\} 
= \frac{1}{\sqrt{|g|} g^{00}}\, \delta_{\alpha\beta }\, \delta({\bf x} - {\bf x}^{\prime}) \,.
\label{cofer}
\end{equation} 
After elimination of the  longitudinal gauge field components by the unitary  transformation (\ref{ugft0}) with the charge density given in Eq.\,(\ref{j0sp}) the final form of the Hamiltonian reads
\begin{equation}
H^{\prime}  = H_A^{\prime} +\int d^{\,d} x \sqrt{|g|} \, \bar{\psi}\big[\gamma^{k}i(\nabla_{k} -i e \hat{A}_{k}) +m + i g^{00}\, \Gamma_0\big]\psi\,, 
\label{hamf}
\end{equation}
and the Gau\ss\ law constraint (\ref{gp}) insures the vanishing of the longitudinal component of the conjugate momentum in the space of physical states.
\subsection{SU(N) Yang-Mills fields in Weyl gauge}
The first step in the above procedure is straightforwardly generalized to Yang-Mills gauge fields.
Starting from the action (with the summation convention applied also to the color labels $a,b,c$)
 \begin{equation}
S = \int d^{\,d+1}x \sqrt{|g|}\Big[- \frac{1}{4} g^{\mu \rho} g^{\nu \sigma} F_{\mu\nu}^a F_{\rho\sigma}^{a}\Big]\,,
\label{laym}
\end{equation}
with the field strength components 
\begin{equation} F_{\mu\nu}^a = \partial_{\mu} A_{\nu}^a - \partial_{\nu} A_{\mu}^a +g_{YM} f^{abc}A_{\mu}^bA_{\nu}^c, \quad a=1,\ldots, N^2-1 
\label{ymfs}
\end{equation}
one easily verifies that in the Weyl gauge 
$$A_0^a=0\,,$$
the formal expression for the conjugate momentum (\ref{com}) remains unchanged 
\begin{equation}
\label{comym}
\Pi^{a\,i} = \frac{\partial {\cal L}}{\partial \partial_{0} A_{i}^a} = - \sqrt{|g|}\, g^{0 0}\, \partial_{0} A^{a\,i} \, ,
\end{equation}
and so do the  commutation relations 
$$[\Pi^{a\,i}(x^0, {\bf x}), A_{j}^b(x^0,{\bf x}^{\prime})] 
= \frac{1}{i}\delta_{a b}\delta_{i j } \delta({\bf x} - {\bf x}^{\prime}) \,. $$
Also the structure of the Hamiltonian is the same as in (\ref{ham2}) 
\begin{equation}
H = \int d^{\,d} x  \Big\{-\frac{1}{2} \frac{g_{i j}}{\sqrt{|g|} \,g^{00}} \Pi^{a\,i}\Pi^{a\,j} + V[A]  \Big\}  \, ,  \label{hamym}
\end{equation}
with the magnetic field energy given by
\begin{eqnarray}
 V[A] &=& \frac{1}{4} \sqrt{|g|}  g^{ik} g^{jl} F_{ij}^a F_{kl}^a
 \, .
 \label{magfiym}
\end{eqnarray}
The  Gau\ss\ law is  implemented as constraints on the physical states 
\begin{equation}
G^a |\Psi_p\rangle =0\,, \quad  G^a = \Big(\partial_i \delta_{ac} + g_{YM}f^{abc}A^b_i \Big)\Pi^{c\,i} \,.
\label{glym}
\end{equation}
As in electrodynamics, the components of the Gau\ss\ law operator commute with the Hamiltonian
$$ [G^a , H] =0\,,$$
and satisfy, as in flat space, the commutation relations 
\begin{equation}
[G^a({\bf x}),G^b({\bf y})]= ig_{YM} f^{abc} G^c({\bf x}) \delta({\bf x}- {\bf y})\,.
\label{coga}
\end{equation}   
Together with the commutation relations (\ref{coga}), the vanishing commutator of Hamiltonian and Gau\ss\ law operator indicates, as in the Maxwell case,  the possibility of  eliminating   degrees of freedom. In Yang-Mills theory, the appearance  of Gribov horizons prevents  an explicit construction of  the Hamiltonian in the Coulomb gauge by elimination of the longitudinal fields already in flat space. Depending on the structure of space-time axial gauges may be more appropriate. The Rindler space  or the AdS metric, for instance,  singles out one space direction and no further reduction of the symmetry occurs if the axial gauge is defined with respect to this direction. With the technique of elimination of redundant degrees of freedom by unitary transformations the Gauss law constraint (\ref{glym}) can be fully implemented \cite{LNT94,LOY07} and the gauge field component with respect to the particular space direction disappears.
\section{Scalar fields in Rindler space}
\subsection{The wave equation}
In order to prepare our discussion of gauge fields,  we sketch in this section the treatment of scalar fields in Rindler space and its connection to scalar fields in Minkowski space. We follow here closely the treatment given in \cite{FULL73}. 
We consider a scalar field  in a frame which is uniformly accelerated in the 1-direction with respect to an inertial frame (coordinates $(t, x^1)$).
Coordinates  in the accelerated system can be chosen to be related to the coordinates $(t, x^1)$ in the right Rindler wedge
\begin{equation}
R_+ = \big\{x^\mu\big|\,|t|\le x^1 \big\}\,                           
\label{rau3}
\end{equation}
in the following way 
\begin{equation}
t (\tau, \xi) = \frac{1}{a} \; e^{a \xi} \sinh a \tau\,, \quad 
x^1 (\tau, \xi) = \frac{1}{a} \; e^{a \xi} \cosh a \tau \, .          \label{cd3}
\end{equation}
The range of $\tau,\,\xi $ is 
\begin{equation}
- \infty \; \le \; \tau\,,\xi \; \le \; \infty .                               \label{rau2}
\end{equation}
The metric of the Rindler space-time with $d-1$ transverse directions  (${\bf x}_{\perp}$) is given by 
\begin{equation}
ds^{2} = e^{2 a \xi} (d \tau^{2} - d \xi^{2})- d{\bf x}^2_{\perp} \,.             
\label{rin}
\end{equation}
The action of a self interacting scalar field reads
\begin{equation}
S = \frac{1}{2} \int d \tau \; d \xi \; d^{\,d-1}x_{\perp} \big \{ (\partial_{\tau} \phi )^{2}
- (\partial_{\xi} \phi )^{2} - (m^{2} \phi^{2} + (\boldsymbol{\nabla}_{\perp} \phi)^{2}-2 V(\phi) ) \;
 e^{2 a \xi} \big \} \, .                                                           \label{ac}
\end{equation}
The equation of motion for a non-interacting scalar field 
\begin{equation}
\Big [ \partial^{2}_{\tau} - \Delta_s + m^{2} \;
e^{2 a \xi} \Big ]  \; \phi = 0   \, ,                              \label{EQM1}
\end{equation}
with the Laplacian
\begin{equation}
\label{scala}
\Delta_s = \partial^{2}_{\xi} +e^{2 a \xi}\boldsymbol{\nabla}_{\perp}^2\, , 
\end{equation}
is solved with the ansatz 
\begin{displaymath}
\phi = e^{- i \omega \tau} \; e^{i {\bf k} _{\perp} {\bf x}_{\perp}} \; \varphi (\xi) \, .
\end{displaymath} 
 With 
\begin{equation}
 m^{2}_{\perp}= m^{2} + {\bf k}^{2}_{\perp}\,,
\label{mperp}
\end{equation}
the wave equation  for $\varphi$ reads
\begin{equation}
- \frac{d^{2} \varphi}{d \xi^{2}} + m^{2}_{\perp} \; e^{2 a \xi} \varphi = \omega^{2} \varphi \,.     \label{wxi}
\end{equation}
In terms of the variable 
\begin{equation}
z=\frac{m_{\perp}}{a } e^{a \xi}  \, , 
\label{z}
\end{equation}
the eigenfunctions 
\begin{equation}
\varphi (z) = k_{i \frac{\omega}{a}} (z) \,                 \label{kio}
\end{equation}
are  (appropriately normalized) MacDonald functions 
\begin{equation}
k_{i \mu} (z) = \frac{1}{\pi} \; \sqrt{2 \mu \sinh  \pi \mu } \; K_{i \mu} (z)  ,\quad     \label{kmu}
\end{equation}
which form a complete orthonormal  set
\begin{equation}
\int^{\infty}_{0} \frac{dz}{z} k_{i \mu} (z) \; k_{i \nu} (z) = \delta (\mu - \nu) ,\quad  \int^{\infty}_{0} d \mu \; k_{i \mu} (z) \; k_{i \mu} (z^{\prime}) = z \delta (z-z^{\prime})  \,.     \label{ort}
\end{equation}
\subsection{Hamiltonian and scale transformations}\label{hmst}
%\label{unhas}
 According to the action  (\ref{ac}), the momentum conjugate to the scalar field $\phi$ is 
\begin{equation}
\pi = \partial_{\tau} \phi \, ,
\label{como}
\end{equation}
and the Hamiltonian of a free scalar field  in the accelerated system reads 
\begin{equation}
H_0^{sc} = \frac{1}{2} \int d \xi \, d^{\,d-1}x_{\perp} \big \{  \pi^{2} + (\partial_{\xi}\phi)^{2}
+ (m^{2} \phi^{2} + (\boldsymbol{\nabla}_{\perp}\phi)^{2})\, e^{2 a \xi} \big \} \,.           \label{han}
\end{equation}
We quantize the scalar field by requiring canonical commutation relations
\begin{equation}
  \label{cacos}
  [\pi (\tau, \xi,{\bf x}_{\perp} ) , \phi (\tau,\xi^{\prime}, {\bf x}_{\perp}^{\prime}) ] =\frac{1}{i}\, \delta (\xi - \xi^{\prime})\,\delta ({\bf x}_{\perp}-{\bf x}_{\perp}^{\prime})\, .
\end{equation}
The  creation and annihilation operators defined by the normal-mode expansion of $\phi$ 
\begin{equation}
\phi (\tau, \xi, {\bf x}_{\perp}) = \int \frac{d \omega}{\sqrt{2\omega}} \frac{d^{\,d-1}k_{\perp}}{(2\pi)^{(d-1)/2}} (a (\omega, {\bf k}_{\perp}) e^{- i \omega \tau + i {\bf k} _{\perp} {\bf x}_{\perp}}
+ a^{\dagger}(\omega, {\bf k}_{\perp}) e^{i \omega \tau - i {\bf k}_{\perp} {\bf x}_{\perp}})\,
\, k_{i \frac{\omega}{a}} (z)   
\label{nomosc}
\end{equation}
satisfy standard commutation relations
\begin{equation}
[ a (\omega, {\bf k}_{\perp}), a^{\dagger} (\omega^{\prime}, {\bf k}_{\perp}^{\prime}) ] = \delta (\omega - \omega^{\prime})  \delta ( {\bf k}_{\perp} -  {\bf k}_{\perp}^{\prime})\,.
\label{crta}
\end{equation}
With the help of the orthogonality relation (\ref{ort}) the Hamiltonian  (\ref{han})  can be expressed in terms 
of creation and annihilation operators
\begin{equation}
H_0^{sc} = \int d^{\,d-1}k_{\perp} \int^{\infty}_{0} d \omega\, \omega  a^{\dagger} (\omega, {\bf k}_{\perp}) a (\omega, {\bf k}_{\perp})
 \,  .
\label{hamsca}
\end{equation}
The energies of the elementary excitations, the ``Rindler particles'', are independent of the transverse momentum, giving rise to an infinite degeneracy of all eigenstates of $H_0^{sc}$ including the vacuum. This degeneracy is a consequence of a symmetry. We consider first the case of a massless field. For $m=0$ and vanishing $V(\phi)$ the action (\ref{ac}) is invariant under the transformation
\begin{equation}
  \label{ptp}
\phi (\tau, \xi, {\bf x}_{\perp}) \longrightarrow e^{(d-1)a \xi_{0}/2}\, \phi (\tau, \xi^{\prime},  {\bf x}^{\prime}_{\perp})\,,
\end{equation}
consisting of a translation in $\xi$ and a rescaling of the perpendicular coordinates ${\bf x}_{\perp}$
\begin{equation}
  \label{trsc}
\xi^{\prime} = \xi+\xi_0\,,\quad  {\bf x}^{\prime} _{\perp}= e^{a \xi_{0}} {\bf x}_{\perp} \,.
\end{equation}
The limit of infinitesimal transformations determines the generator of the symmetry 
\begin{equation}
\label{QM0}
Q = \int d \xi d^{d-1} x_{\perp} \pi (\xi, {\bf x}_{\perp})  \left \{ \partial_{\xi} + a \left (\frac{d-1}{2} + 
{\bf x}_{\perp} \boldsymbol{\nabla}_{\perp} \right )  \right \} \phi \,.
\end{equation}
The Hamiltonian $H_{0,\,m=0}^{sc}$, the components $i\neq 1$ of the momentum operator 
\begin{equation}
P_{i} = i \int d \xi d^{d-1} x_{\perp} \; \pi (\xi, {\bf x}_{\perp}) \partial_{i} \phi (\xi,  {\bf x}_{\perp}) 
\end{equation}
and  the generator $Q$ form the algebra 
\begin{eqnarray}  % zu lang
\big[ H_{0,\,m=0}^{sc}, Q \big] = 0 \,,\quad \big [ P_{i}, Q \big ]   =   - i a P_{i}  \,, \quad \big[P_i,H_{0,\,m=0}^{sc}\big]=0 \quad i\neq 1\,.
\label{algpq}
\end{eqnarray}
Furthermore $ H_{0,\,m=0}^{sc}$ and $Q$ commute with the generator of rotations 
$$M^{ij}= x^iP^j-x^jP^i,\quad \big[ M^{ij}, H_{0,\,m=0}^{sc} \big]= \big[ M^{ij}, Q \big] =0\,, i,j=2,\ldots, d-1\,, $$
and thus by applying rotations and the transformations (\ref{ptp}),   a stationary state with a given perpendicular momentum can be transformed into a stationary state of the same energy and with   arbitrary  perpendicular momentum.  

 By the coordinate change  $x^1= e^{a\xi}/a$, the transformation (\ref{ptp},\ref{trsc}) can be converted into  a scale transformation in all spatial coordinates $x^i$ with the standard form of the commutator between the generator of the scale transformations $Q$ and $P_i$. As a  special property of Rindler space, the transformation (\ref{ptp}) does not affect the time coordinate  and leaves therefore the (dimensionful) Hamiltonian invariant under scale changes. Expressed in terms of the Minkowski-space coordinates (\ref{cd3}), the transformation (\ref{ptp}) becomes the standard scale transformation of the fields in the Rindler wedge of the Minkowski space  including the scaling of the time coordinate. Unlike the Rindler space Hamiltonian, the Minkowski space Hamiltonian changes under these transformations and no degeneracy in the spectrum results.   

The energies of the Rindler particles do not depend either on their mass.  After Fourier transform in the perpendicular coordinates the following transformation of the fields
\begin{equation}
\phi (\tau, \xi, {\bf k}_{\perp}) \longrightarrow \phi (\tau, \xi + \xi_{0}, \; 
\lambda ({\bf k}^{2}_{\perp})\, {\bf k}_{\perp}) \gamma ({\bf k}_{\perp}) \,, \quad
\end{equation}
with 
\begin{equation}
\lambda^{2} ({\bf k}^{2}_{\perp}) = e^{-2 a \xi_{0}} + \frac{m^{2}}{{\bf k}^{2}_{\perp}} \; 
(e^{-2 a \xi_{0}} - 1)\,,\quad \gamma^2 ({\bf k}_{\perp}) = \Big|\det \frac{\partial}{\partial k_{I}} \,\big( \lambda ({\bf k}^{2}_{\perp}) k_{J}\big)\Big|\,,
\end{equation}
leaves the action invariant. The generator of the symmetry is non-local 
\begin{eqnarray*}
Q = \int d \xi d^{d-1} x_{\perp} \pi (\xi, {\bf x}_{\perp})&&
\hspace{-.3cm}
\Bigg \{\partial_{\xi} + a \Big(\frac{d-1}{2} + 
{\bf x}_{\perp} \boldsymbol{\nabla}_{\perp} \Big)   \phi 
\\ &&
\hspace{-.2cm}
+a\,m^{2} \Big (\frac{d-3}{2} + {\bf x}_{\perp} \boldsymbol{\nabla}_{\perp} \Big)\int 
d^{d-1} y_{\perp} g(|{\bf x}_{\perp}-{\bf y}_{\perp}|)\phi (\tau, \xi, {\bf y}_{\perp})\Bigg\}\,.
\end{eqnarray*} 
with the non-locality determined by the static propagator
\begin{displaymath}
 g(x_{\perp})=
\frac{\Gamma(\frac{d-3}{2})}{4\pi^{(d-1)/2}} \frac{1}{x^{d-3}_{\perp}}\,\,.
\end{displaymath}
\subsection{Unruh temperature}\label{unhas}
Up to this point we have discussed  the quantized scalar field with its excitations, the ``Rindler particles'',  as a field theory in  a space with a simple but non-trivial metric. The main interest in studying this field theory is its relation with the corresponding theory in Minkowski space.    
Starting point for connecting the dynamics in inertial and uniformly accelerated systems is the requirement that
 the value of a scalar field is the same  in inertial  ($\tilde{\phi}$)  and accelerated ($\phi$) frames
\begin{equation}
\phi (\tau,\xi,{\bf x}_{\perp}) = \tilde{\phi} (t,{\bf x}) \; \Big |_{t,{\bf x} = t,{\bf x} (\tau, \xi)} \, .
\label{phtph}
\end{equation}
In the inertial frame we have the standard normal-mode expansion
\begin{equation}
\tilde{\phi} (t, {\bf x}) =\int \frac{d^{d}k}{\sqrt{2\omega_k}(2\pi)^{d/2}} (\tilde{a} ({\bf k} ) e^{- i \omega_k t + i {\bf k} {\bf x}}
+ \tilde{a}^{\dagger}({\bf k}) e^{i \omega_k t - i {\bf k} {\bf x}})\, .
\nonumber\\                                                                \label{nomosca}
\end{equation}
By projecting Eq.\,(\ref{phtph}) on the normal modes of the field $\phi$  in the accelerated system (cf.\,\cite{MUWI07}) and with the help of the identity    (cf. \cite{RG65})
\begin{equation}
 \int^{\infty}_{0} \frac{dz}{z} k_{i \frac{\Omega}{a}} (z) e^{i \kappa z} = \sqrt{\frac{2}{\frac{\Omega}{a} \sinh \pi \frac{\Omega}{a}}} \cos \frac{\Omega}{a}
\left (\beta_{\bf k} - a\tau  - \frac{i \pi}{2} \right )                                    \label{kphim}
\end{equation}
with
\begin{equation}
\kappa = \sinh (\beta_{{\bf k}} - a \tau) \, ,\quad \sinh \beta_{\bf k} = \frac{k_1}{m_{\perp}}\, ,
\label{beta}
\end{equation}
the  relation between creation and annihilation operators in the two frames
\begin{equation}
a (\Omega, {{\bf k}}_{\perp}) = \frac{1}{\sqrt{a \sinh \pi \frac{\Omega}{a}}} \int^{\infty}_{- \infty}
\frac{dk_1}{\sqrt{2 \pi 2 \omega_k}}  e^{i \frac{\Omega}{a} \beta_{{\bf k}}} \left [
e^{\frac{\pi \Omega}{2 a}} \tilde{a} (k_1, {\bf k}_{\perp})  + e^{- \frac{\pi \Omega}{2 a} } \tilde{a}^{\dagger}
(k_1, -{{\bf k}}_{\perp}) \right ]\,,                                                   \label{alpa}
\end{equation}
is obtained.
With  this result  the average particle number in the Minkowski space vacuum $ |0_M \rangle$ as seen by the accelerated observer 
is easily computed
\begin{equation}
\langle 0_M | a^{\dagger} (\Omega, {\bf k}_{\perp})  a (\Omega^{\prime}, {\bf k}_{\perp}^{\prime}) | 0_M \rangle
=\frac{1}{e^{2 \pi \frac{\Omega}{a}} - 1} \delta (\Omega - \Omega^{\prime})\delta({\bf k}_{\perp}-{\bf k}^{\prime}_{\perp}) \, .        
\label{urt2}
\end{equation}
The particle number exhibits a thermal distribution. In comparison with  thermal distributions in Minkowski space  the thermal distribution  of ``Rindler particles'' does not suppress states of arbitrarily large  transverse momentum  ${\bf k}_{\perp} $. Furthermore there is no degeneracy related to the  motion in the direction of the acceleration.  There are no modes of excitations incoming from positive $\xi $. For various applications,  regularization of the infinities related to the degeneracy of the energy with respect to  the transverse momenta is required and the detailed form of the resulting  distribution will depend on this regularization of the ultraviolet momenta. The regularization may be achieved by introducing a sharp ultraviolet cutoff as used e.g. in the calculation \cite{SULE05} of the entropy of a scalar field in Rindler space.  An effective regularization \cite{CADE77,BIDI82,SAHA02}, affecting also the infrared properties, is obtained by modifying the Rindler space and introducing an infinite barrier which confines the system to $\xi > 0$. In Minkowski space this corresponds to an accelerated ``mirror''.  Such boundary conditions  break the scale invariance of the action (cf.\,Eq.\,(\ref{ptp})) and therefore lift the degeneracy of the eigenstates of the Hamiltonian. For instance,  the requirement for the eigenfunctions $\varphi$ (\ref{kio}) to vanish at $\xi = 0$ can be satisfied only if  (for $m=0$) this point is in the classically allowed regime of the Schr\"odinger-like equation (\ref{wxi}). This implies roughly 
$$\omega^2 \ge {\bf k}_{\perp}^2 . $$  
Thus the boundary condition results in a connection between the energy and the transverse momentum. With this regularization,  and for $\xi\gg a^{-1}$,  the local distribution becomes Planckian  \cite{CADE77} provided the ``improved'' energy momentum tensor is used.\\
As has been shown in \cite{UNRU76}  (cf.\,\cite{UNWA84},\,\cite{CRHM07}) the thermal distribution arises due to the  elimination of the degrees of freedom in the left Rindler wedge 
\begin{equation}
R_- = \big\{x^\mu\big|\, x^1 \le -|t| \big\}\, .                          
\label{rau4}
\end{equation}
After this elimination, the dynamics in the right Rindler wedge  is described by the density matrix 
\begin{equation}
  \label{denma}
  \rho = \rho_0 e^{ - 2\pi H_0^{sc}/a} \,, \quad \rho_0^{-1} = \mbox{tr}\,  e^{ - 2\pi H_0^{sc}/a}\,,  
\end{equation}
with the Rindler space Hamiltonian (\ref{hamsca}).
\section{Electromagnetic  fields in Rindler space}
\subsection{Canonical formalism}\label{cafo}
Here we apply the canonical formalism developed above to the Maxwell field in Rindler space-time. The Rindler metric (\ref{rin}) 
singles out  the 1-direction  and in the following we will refer  to  the $d$ spatial components (with $x^1=\xi$)  with small  and   to the  $d-1$ perpendicular components with capital Latin letters. 
The simplicity of the metric (\ref{rin}) gives rise to significant simplifications in the expressions of the relevant operators of the quantized theory. In particular, with $$\sqrt{|g|} g^{00}=1\,,$$
it belongs to the class of metrics (\ref{rili}) where static charges do not couple to the transverse electric field. In the transformation (cf.\,Eq.\,(\ref{dxxp2})) to the frequently used coordinate $x^{\prime\,1}=e^{a x^{1}}/a$ (cf. \cite{FULL73},\cite{DAVI75})  this simplifying feature gets lost and charge densities either  external or associated with matter fields  couple to the radiation field. On the other hand, this coordinate transformation makes the construction of longitudinal and transverse gauge fields identical to that in Minkowski space.    We will now discuss the details of the quantized   non-interacting Maxwell field in Rindler space using the metric (\ref{rin}).  As shown in Sec.\,\ref{sec:decpl} the coupling to external charges can be accounted for by including the classical solutions to the corresponding inhomogeneous equations of motion (\ref{GLC0}) and (\ref{eqch0}) and will be addressed in Sec.\,\ref{sec:riex}. 
 In Rindler space, the explicit form of the action of the Maxwell field (cf.\,Eq.\,(\ref{stame})) reads
\begin{eqnarray}
S & = & \frac{1}{2} \int d \tau d \xi d^{d-1} x_{\perp} \Big \{ e^{- 2 a \xi} ( \partial_{0} A_{1} -
\partial_{1} A_{0})^{2} + \sum^{d}_{I = 2} \Big [ ( \partial_{0} A_{I} - \partial_{I} A_{0})^{2} \nonumber  \\
& & - (\partial_{1} A_{I} - \partial_{I} A_{1} )^{2} \Big ] - e^{2 a \xi} \sum^{d}_{J > I = 2}
(\partial_{I} A_{J} - \partial_{J} A_{I})^{2} \Big \} \,,
\label{sedr}
\end{eqnarray}
and, in terms of transverse fields,  the Hamiltonian (\ref{hp2}) is given by 
\begin{equation}
H_0 = \frac{1}{2} \int  d \xi d^{\,d-1}x_{\perp} \Big\{ e^{2a\xi} \hat{\Pi}^{1\,2}+ \sum_{I=2}^d \hat{\Pi}^{I\,2} 
+ \sum_{I=2}^d \big( F_{1I}^2 + e^{2a\xi}\sum_{J>I}^d  F_{IJ}^2 \big)\Big\}\,,
\label{hapr3}
\end{equation}
with 
$$ F_{ij}=\partial_i\hat{A}_j -\partial_j\hat{A}_i\,. $$
To complete this explicit formulation of the dynamics we have calculated the canonical commutators (\ref{patr}) and sketch here the calculation for $i=j=1$. 
Using  the identity  (cf.\,Eq.\,(\ref{lapla}))
\begin{equation}
  \label{d11}
1 + \partial^{1} \; \frac{1}{\Delta} \; \partial_{1} =  \frac{1}{\Delta_s}\;
e^{2 a \xi} \partial^{2}_{i}\, ,
\end{equation}
and the completeness relation of the eigenfunctions of the Laplacian  $\Delta_s$  of the scalar wave equation (\ref{scala})  (cf.\,Eq.\,(\ref{ort}))
we obtain
\begin{equation}
[ \hat{\Pi}^{1} (\tau,x^{\prime}), \hat{A}_{1} (\tau,x)] = -i \int \frac{d^{\,d-1}k_{\perp}}{(2 \pi)^{\,d-1}} \; z^2\,
e^{i {\bf k}_{\perp}({\bf x}_{\perp} - {\bf x}^{\prime}_{\perp})} \int d \,\omega \; \frac{a^2}{\omega^{2}} \;
k_{i \frac{\omega}{a}} (z^{\prime}) k_{i \frac{\omega}{a}} (z)
 \,,                          
\label{com11}
\end{equation}
with the  variable $z$  defined by (cf.\,Eq. (\ref{z}))
\begin{equation}
z=\frac{k_{\perp}}{a} e^{a\xi}\,.
\label{zkp}
\end{equation}
This integral, and similar ones for the other commutators,  can be  evaluated in closed form. For $d=3$ for instance, we obtain
\begin{equation}
\label{com3}
[ \hat{\Pi}^{1} (\tau,x^{\prime}), \hat{A}_{1} (\tau,x)] = \frac{i}{2 \pi a} e^{2 a \xi}\boldsymbol{\nabla}^{2}_{\perp} \left(s^4+4 s^2 e^{a(\xi+\xi^\prime )}/a^2\right)^{-1/2}\,
\end{equation}
with the geodesic distance
\begin{equation}
s^2 = \frac{1}{a^2}\Big(e^{a\xi}-e^{a\xi^{\prime}}\Big)^2+({\bf x}_{\perp}-{\bf x}_{\perp}^{\prime})^2\, .
\label{sgd}
\end{equation}
 Using similar  identities, the commutator of the other components of the transverse gauge fields and the conjugate momenta can be calculated. 
\subsection{Normal modes of transverse gauge fields}\label{sec:trgf}
The equation of motion of the transverse fields are obtained by transverse projection of the  equations of motion (\ref{coeq}) (or more easily from Eq.\,(\ref{eqri}) by the substitution $\hat{\alpha}^i \rightarrow \hat{A}^i$)
\begin{equation}
\partial_{\tau}^{2}\hat{A}^{1} -\Delta_s \hat{A}^{1} = 0\, ,
\label{notr1}
\end{equation}
\begin{equation}
\partial_{\tau}^{2}\hat{A}^{I} -\Delta_s \hat{A}^{I} -2ae^{2a\xi}\partial^I\hat{A}^{1}= 0\, ,
\label{notri}
\end{equation}
with  the Laplacian $\Delta_s$ of the scalar wave equation (\ref{scala}). 
With the ansatz
\begin{equation}
\hat{A}^{I} = \hat{a}^{I} +  \; \frac{1}{\boldsymbol{\nabla}_{\perp}^2} \; \partial^{I} \partial_{1} \hat{A}^{1} \,,
\label{ansi}
\end{equation}
the equations of motion (\ref{notri}) of the perpendicular  components $\hat{A}^I$ of the transverse gauge fields 
turn  into the wave equation of a scalar field 
\begin{equation}
\partial_{\tau}^{2}\hat{a}^{I} - \Delta_s \hat{a}^{I} = 0 \,  .
\label{wveqi}
\end{equation}
The  normal mode expansion of $\hat{A}^1$ satisfying the equation of motion (\ref{notr1}) for a vanishing current is essentially that of  a scalar field  (Eq.\,(\ref{nomosc}))
\begin{equation}
\hat{A}_{1}(\tau,\xi,{\bf x}_{\perp})=\int \frac{d \omega}{\sqrt{2\omega}} \frac{d^{\,d-1}k_{\perp}}{(2\pi)^{(d-1)/2}}\frac{a^2}{\omega  k_{\perp}} \Big[a_1 (\omega, {\bf k}_{\perp}) e^{- i \omega \tau + i {\bf k} _{\perp} {\bf x}_{\perp}}
+ \text{h.c.}\Big]\,
z^{2}\, k_{i \frac{\omega}{a}} (z)\, .
\label{notr1a}
\end{equation}
To derive the normal mode expansion for the perpendicular components we observe that the  transversality condition (cf.\,Eqs.\,(\ref{proj3},\ref{traug})) applied to Eq.(\ref{ansi}) yields
\begin{equation}
\partial_{I} \hat{a}^{I}=0 \, ,
\end{equation}
which, as in flat space can be implemented by defining creation and annihilation operators with respect to a transverse basis $e^{L}_{i} ({\bf k}_{\perp}),\quad L = 2,\ldots,d-1 \,  $  of polarization vectors with the properties
\begin{equation}
\sum_{I=2}^{d} k_{I} e^{L}_{I} ({\bf k}_{\perp} ) = 0 \,, \quad 
\sum_{I=2}^{d} e^{L}_{I} ({\bf k}_{\perp}) e^{L^{\prime}}_{I} ({\bf k}_{\perp}) = \delta_{L,L^{\prime}}\,,\quad \sum^{\,d-1}_{L = 2} e^L_{I} ({\bf k}_{\perp}) \; e^L_{J}  ({\bf k}_{\perp}) 
= \delta_{IJ} - \frac{k_{I} k_{J}}{{\bf k}_{\perp}^{2}}\,  .
\label{pol}
\end{equation}
The normal mode expansion  
\begin{eqnarray}
&&\hat{A}_{I}(\tau,\xi,{\bf x}_{\perp})=\int \frac{d \omega}{\sqrt{2\omega}} \frac{d^{\,d-1}k_{\perp}}{(2\pi)^{(d-1)/2}}\nonumber\\
&&\cdot\Big[ \Big\{\sum_{L=2}^{\,d-1} e^{L}_{I} ({\bf k}_{\perp})
a_L (\omega, {\bf k}_{\perp}) + i \frac{a\,k_{I}}{\omega k_{\perp}} a_{1}(\omega, {\bf k}_{\perp}) z \frac{d}{dz}\Big\} k_{i \frac{\omega}{a}}(z)
e^{- i \omega \tau + i {\bf k}_{\perp}{\bf x}_{\perp}}+\text{h.c.}\Big]  \, ,\label{trnoi}
\end{eqnarray}
is thereby expressed in terms of the $d-2$ unconstrained amplitudes  $a_L,\, L=2, d-1$.
With the above choice of the  normalization in Eqs.\,(\ref{notr1a}) and (\ref{trnoi}),  and requiring  standard commutation relations for creation and annihilation operators
\begin{equation}
 [a_{i}(\omega,{\bf k}_{\perp}), a_{j}^{\dagger}(\omega^{\prime},{\bf k}_{\perp}^{\prime})]=  \delta_{ij}\,\delta (\omega - \omega^{\prime})\,\delta({\bf k}_{\perp}-{\bf k}_{\perp}^{\prime})\,, \quad i,j=1,\ldots,d-1\, ,
\label{cocran}
\end{equation}
the commutation relations of the transverse gauge fields (cf. Eq.(\ref{com11})) as implied by the canonical commutation relations (\ref{caco}) and the definition of the transverse gauge fields (\ref{traug}) are satisfied. \\
\subsection{Hamiltonian and scale transformations}\label{EDHS}
In terms of the creation and annihilation operators, the Hamiltonian is given by 
\begin{equation}
H_0 =  \int d^{\,d-1}k_{\perp} \int d \omega  \omega \sum_{i=1}^{\,d-1} a^{\dagger}_{i}
(\omega, {\bf k}_{\perp}) a_{i} (\omega,  {\bf k}_{\perp})  \, .                           \label{Eng2}
\end{equation}
The same degeneracy with respect to the transverse momenta occurs as for the scalar field (cf.\,Eq.\,(\ref{hamsca})) and can be analyzed  following the discussion  in Sec.\,\ref{hmst}. The  transformation of the Maxwell field leaving the action (\ref{sedr})  invariant is given by (cf.\,Eqs.\,(\ref{ptp}), (\ref{trsc})) 
\begin{eqnarray}
A_{1,0} (\tau, \xi, {\bf x}_{\perp}) &\to& e^{(d-3)a \xi_{0}/2} A_{1,0} (\tau, \xi^{\prime}, {\bf x}_{\perp}^{\prime})\,, 
\nonumber \\
A_{I} (\tau, \xi, {\bf x}_{\perp}) &\to& e^{(d-1) a \xi_{0}/2} A_{I} (\tau, \xi^{\prime}, {\bf x}_{\perp}^{\prime})\,. 
\label{sctA} 
\end{eqnarray}
The perpendicular components $A_{I}$ carry the same weight as the scalar field (\ref{ptp}).
The identity 
$$\partial_i A^i(\tau,\xi,{\bf x}_{\perp}) \to e^{(d+1)a\xi_0/2}\partial^{\prime}_i A^i(\tau, \xi^{\prime}, {\bf x}_{\perp}^{\prime})$$
guarantees that  under the symmetry transformation, longitudinal and transverse components of the gauge field do not mix. The generator of the symmetry transformations,
\begin{displaymath}
Q =\int d \xi d x_{\perp} \Bigg\{ \sum^{d}_{i = 1} \Pi^{i} \Big[\partial_{\xi} + a\big(\frac{d-1}{2}+{\bf x}_{\perp}\boldsymbol{\nabla}_{\perp}\big)\Big]A_{i} -a \,\Pi^1 A_1\Bigg\} \,,
\end{displaymath}
the Hamiltonian and the perpendicular components of the  momentum operator 
\begin{equation*}
P_{I} = i \int d \xi d^{d-1} x_{\perp} \; \sum_{k=1}^{d}\Pi^k (\tau,\xi, {\bf x}_{\perp}) \partial_{I}A_k (\tau,\xi,  {\bf x}_{\perp}) 
\end{equation*}
satisfy the commutation-relations (\ref{algpq}). Together with the rotations in the perpendicular coordinates, these transformations are responsible for the degeneracy of the stationary states.  

The basis for extending   the symmetry considerations to interacting gauge fields is the behavior of the (gauge-) covariant derivative under the scale transformations (Eqs.\,(\ref{sctA}), and (\ref{trsc})) 
\begin{eqnarray}
\partial_1-i e A_1(\tau,\xi,{\bf x}_{\perp}) &\to& \partial_1^{\prime} -i \hat{e}(\xi_0)\, A_1 (\tau, \xi^{\prime}, {\bf x}_{\perp}^{\prime})\nonumber\\
\partial_I-i e A_I(\tau,\xi,{\bf x}_{\perp}) &\to& e^{a\xi_0} \big(\partial_I^{\prime} -i\hat{e}(\xi_0)\, A_I (\tau, \xi^{\prime}, {\bf x}_{\perp}^{\prime})\big)\,,
\label{covde}
\end{eqnarray}
with the ``running coupling constant'' 
\begin{equation}
\hat{e}(\xi_0) = e^{(d-3)a\xi_0/2} e\,.
\label{ruco}
\end{equation}
Under the combined transformation of the scalar  (\ref{ptp}) and the gauge field (\ref{sctA}) the action of scalar QED with a massless scalar field is thus reproduced up to a change in the coupling constant
$$S[A,\phi,e] \to S[A,\phi,\hat{e}(\xi_0)]$$\.
 In the   4-dimensional Rindler space ($d=3$), the combined transformation of scalar and gauge fields leaves the action invariant. The same argument applies for massless fermions coupled to gauge fields with action (cf. \cite{BIDI82}, \cite{ORIT00})
$$S= \int d \tau d \xi d^{d-1} x_{\perp} \bar{\psi}\, i\Big \{\gamma^{0}\big(\partial_{0} -ieA_{0}\big) +\gamma^{1}\big(\partial_{1} +\frac{a}{2}-ieA_{1}\big) +e^{a\xi}\gamma^I \big(\partial_I -ie A_I\big)\Big\}\psi\,,$$ 
with $\gamma^{\mu}$ denoting here the flat space Dirac matrices.
The transformation of the fermion fields
 \begin{equation}\psi(\tau,\xi, {\bf x}_{\perp}) \to e^{(d-1)a \xi_{0}/2} \psi (\tau, \xi^{\prime}, {\bf x}^{\prime}_{\perp})\,,
\label{trff}
\end{equation}   
combined with the transformation of the gauge fields yields
 \begin{equation}
 S[A,\psi,e] \to S[A,\psi,\hat{e}(\xi_0)]\,.
\label{srco}
\end{equation}
Finally we may apply this reasoning  to $SU(N)$ gauge theories and use  the transformation  (\ref{sctA}) for each color component of the gauge field with  the resulting transformations  of the field strength components (\ref{ymfs})
 \begin{equation} F^a_{01}(\tau,\xi,{\bf x}_{\perp}, g_{YM}) \to  e^{(d-3)a\xi_0/2} F^a_{01}(\tau^{\prime},\xi^{\prime},{\bf x}^{\prime}_{\perp}, \hat{g}_{YM}(\xi_0))\,, 
\label{fistr}
\end{equation}
with 
$$\hat{g}_{YM}(\xi_0) = e^{(d-3)a\xi_0}\,g_{YM}\,.$$ 
The  transformation of the other components of the field strength  $F^a_{0I},F^a_{1I},F^a_{IJ}$ differ in  their  weights, with   $e^{(d-3)a\xi_0/2}$ in (\ref{fistr}) being replaced by $e^{(d-1)a\xi_0/2},e^{(d-1)a\xi_0/2},e^{(d+1)a\xi_0/2}$ respectively. Similarly if massless fermion fields are minimally coupled to the gauge fields each of their color components $\psi^{\alpha}$ is subjected to the transformation (\ref{trff}) with the  total change in the action given by  
\begin{equation}
 S[A,\psi,g_{YM}] \to S[A,\psi,\hat{g}_{YM}(\xi_0)]\,.
\label{YMTR}
\end{equation}
Thus, in 4-dimensional Rindler space, the combined transformation (\ref{fistr},\ref{trff}) leaves  the action of QCD (with massless quarks) invariant.   
\subsection{Unruh temperature of the radiation field}
Following  the method outlined in Sec.\,\ref{unhas} we derive the relation of the gauge field operators between inertial and accelerated frames. 
We denote with $\tilde{A}_{i}(t,x,{\bf x}_{\perp})$ the gauge field in the inertial system in the Weyl gauge 
\begin{equation*}
\tilde{A}_{0}(t,x,{\bf x}_{\perp})= 0.
\end{equation*}
Under the coordinate transformation (\ref{cd3}) the $i\ge 2$ components of the gauge field remain invariant, while 
\begin{equation}
\tilde{A}^{\prime\prime}_{0}= e^{\xi a}  \tilde{A}_{1} \sinh a \tau \,,\quad 
\tilde{A}^{\prime\prime}_{1} = e^{\xi a} \tilde{A}_{1} \cosh a \tau\, .
\label{cta}
\end{equation}
The transformed field $\tilde{A}^{\prime\prime}$ can be identified with the gauge field $A$ defined in the accelerated system 
only up to a gauge transformation.  In particular for the comparison with $A$ we have to transform $\tilde{A}^{\prime\prime}$ to the Weyl gauge.  
This is achieved by the gauge transformation
\begin{equation}
\tilde{A}^{\prime}_{\mu}=\tilde{A}^{\prime\prime}_{\mu}- \partial_{\mu}\chi(\tau, \xi, {\bf x}_{\perp})\,, 
\label{gt1}
\end{equation}
with  $\partial_{\mu}$ denoting the derivatives with respect to the coordinates   $\tau, \xi, {\bf x}_{\perp}$. 
The gauge function 
\begin{equation}
\chi (\tau,\xi, {\bf x}_{\perp}) = \int^{\tau} d \tau^{\prime} \tilde{A}^{\prime\prime}_{0} (\tau,\xi, {\bf x}_{\perp})
 +\chi_{(0)}(\xi, {\bf x}_{\perp})
\label{gt2}
\end{equation}
is determined only up to a $\tau-$independent arbitrary function  $\chi_{(0)}$ reflecting the residual gauge invariance 
 in the Weyl gauge.  At this level the transformed fields $\tilde{A}^{\prime}$  can be identified with the gauge fields $A$
\begin{equation}
A_{i}(\tau,\xi,{\bf x}_{\perp}) = 
\tilde{A}_{i}^{\prime} 
(t,{\bf x}) \; \Big |_{t,{\bf x} = t,{\bf x} (\tau, \xi)} \,.
\label{aphtph}
\end{equation}
For translating  this identity  into a relation between the creation and annihilation operators, we have to eliminate the 
longitudinal components. The Gau\ss\ law  
requires  in the space of physical states (with $\tilde{\partial}$ denoting the partial derivatives with respect to $t,{\bf x}$)   
$$\tilde{\Pi}_{(\ell)}^{i}=-\tilde{\partial}_{0} \tilde{A}_{(\ell)}^{i}=0\,,$$
i.e.  the longitudinal gauge field is time independent and can therefore be written as
\begin{equation}
\tilde{A}_{(\ell)}^{\mu}=\tilde{\partial}_{\mu} \tilde{\lambda}(x,{\bf x}_{\perp})\, .
\label{along}
\end{equation}
The transverse field satisfies the covariant transversality condition
\begin{displaymath}
\tilde{\partial}_{\mu} \hat{\tilde{A}}^{\mu} = 0 .
\end{displaymath}
We write the components of the gauge field after the coordinate transformation to the accelerated frame  (\ref{cta}) as 
\begin{equation}
\tilde{A}^{\prime \prime}_{\mu} = \hat{\tilde{A}}^{\prime \prime}_{\mu} 
+ \tilde{A}_{(\ell)\, \mu}^{\prime \prime},  \quad
\tilde{A}^{\prime \prime}_{(\ell) \mu} = \partial_{\mu} \tilde{\lambda}\,,
\label{Lorenz}
\end{equation}
where $\hat{\tilde{A}}^{\prime \prime}_{\mu} $ denotes the transformed components of the  part of the gauge field which is transverse in the inertial frame. 
After  gauge transformation to the Weyl gauge (\ref{gt1})
\begin{eqnarray}
\tilde{A}^{\prime}_{\mu} & = &  \tilde{A}^{\prime \prime}_{\mu} - \partial_{\mu} \; 
\frac{1}{\partial_{0}} \; \tilde{A}^{\prime \prime}_{0} - \partial_{\mu} \chi_{(0)}  (\xi, {\bf x}_{\perp}) \nonumber \\
& = & \hat{\tilde{A}}_{\mu}^{ \prime \prime} + \partial_{\mu} \tilde{\lambda} 
- \partial_{\mu} \; \frac{1}{\partial_{0}} \; \big ( \hat{\tilde{A}}_{0}^{ \prime \prime}
+ \partial_{0} \tilde{\lambda} \big ) - \partial _{\mu} \chi_{(0)}  (\xi, {\bf x}_{\perp}) 
\label{trans}
\end{eqnarray}
we arrive at
\begin{equation}
\tilde{A}^{\prime}_{\mu} = \hat{\tilde{A}}^{\prime \prime}_{\mu} - \partial_{\mu} \; 
\frac{1}{\partial_{0}} \; \hat{\tilde{A}}^{\prime \prime}_{0} - \partial_{\mu} \chi_{(0)} (\xi, {\bf x}_{\perp})\, .
\label{trans2} 
\end{equation}
Here the term $\partial_{\mu} \chi_{(0)} (\xi, {\bf x}_{\perp})$ appears as an yet undetermined, possible  zero mode when inverting $\partial_0$. 
According to Eq.\, (\ref{trans2}), and after adjusting properly this   $\tau$ independent gradient (i.e. subtracting the $\tau$-independent longitudinal field in the accelerated frame),  the transverse field in the accelerated frame is given in terms of the transverse field in the inertial system. We therefore identify the transverse components of the fields appearing in Eq.\,(\ref{aphtph})
\begin{equation}
\hat{A}_{i}(\tau,\xi,{\bf x}_{\perp}) = 
\hat{\tilde{A}}_{i}^{\prime} 
(t,{\bf x}) \; \Big |_{t,{\bf x} = t,{\bf x} (\tau, \xi)} \, .
\label{aphtph2}
\end{equation} 
The explicit expressions for the transverse gauge fields transformed from the inertial to the accelerated frame are
\begin{equation}
\hat{\tilde{A}}_{1}^{\prime} (\xi, \tau, {\bf x}_{\perp})  =  e^{\xi a} \tilde{A}_{1} 
\big (t (\tau, \xi),x (\tau, \xi), {\bf x}_{\perp} \big )-  \partial_{\xi} e^{\xi a} \int^{\tau} d \tau^{\prime} \tilde{A}_{1} ( 
t (\xi, \tau^{\prime}) ,x (\xi, \tau^{\prime}) , {\bf x}_{\perp} )\sinh a \tau^{\prime} , 
\label{ahp1}
\end{equation}
\begin{equation}
\hat{\tilde{A}}_{I}^{\prime} (\xi, \tau, {\bf x}_{\perp})  =  \tilde{A}_{I} 
\big (t (\tau, \xi), x (\tau, \xi), {\bf x}_{\perp} \big )-  \partial_{I} e^{\xi a} \int^{\tau} d \tau^{\prime} \tilde{A}_{1} ( 
t (\xi, \tau^{\prime}) ,x (\xi, \tau^{\prime}) , {\bf x}_{\perp} )\sinh a \tau^{\prime} , 
\label{ahp2}
\end{equation}
with the standard normal mode expansion of the transverse gauge field in the inertial frame
\begin{equation}
\hat{\tilde{A}}_{i}(t,{\bf x})=\int  \frac{d^{d}k}{\sqrt{2\omega_k}\,(2\pi)^{d/2}}
\Big[\sum_{\Lambda=1}^{\,d-1} \tilde{e}^{\Lambda}_{i} ({\bf k})
\, \tilde{a}_{\Lambda} (k_1, {\bf k}_{\perp})
e^{- i \omega_k t + i {\bf k}{\bf x}}+\text{h.c.}\Big]\, . 
\label{aht}
\end{equation}
The $d-1$ dimensional transverse basis $\tilde{\bf e}^{\Lambda}({\bf k})$ is defined in analogy with Eq.\,(\ref{pol}),  and the annihilation $ \tilde{a}_{\Lambda} (k_1, {\bf k}_{\perp})$ and creation $ \tilde{a}_{\Lambda} ^{\dagger}(k_1, {\bf k}_{\perp})$ operators   satisfy standard commutation relations. 

From this point on we may proceed along similar lines as in the case of the scalar field and  project Eq.\,(\ref{aphtph2}) onto the normal modes of the transverse gauge fields in the accelerated system (cf.\,Eqs.\,(\ref{notr1a})\,and\,(\ref{trnoi})). Here we give the details for the $i=1$ component of Eq.\,(\ref{aphtph2}). Projection of  $\hat{A}_{1} $   yields (cf.\,Eq.\,(\ref{zkp}))
\begin{eqnarray}
\int d \xi d^{\,d-1} x_{\perp} \hat{A}_{1} (\tau,\xi, {\bf x}_{\perp}) \frac{1}{z^{2}} \, 
k_{i \frac{\Omega}{a}} (z) e^{- i {\bf k}_{\perp} {\bf x}_{\perp}} 
% \nonumber\\  & & 
=\frac{(2 \pi)^{(d-1)/2} a^2}{\sqrt{2 \Omega} \Omega k_{\perp}} \,  
\Big ( a_{1} (\Omega, {\bf k}_{\perp}) e^{- i \Omega \tau}
+ a^{\dagger}_{1} (\Omega, - {\bf k}_{\perp}) e^{i \Omega \tau} \Big ) \,.
\nonumber  \\ 
&&\label{2}
\end{eqnarray}
Projecting the right-hand side of (\ref{aphtph2})  and using the definition (\ref{ahp1}) we obtain 
\begin{eqnarray}
&&\hspace{-1cm}\int d \xi d^{\,d-1} x_{\perp} \hat{\tilde{A}}_{1}^{\prime} 
(t,{\bf x}) \; \Big |_{t,{\bf x} = t,{\bf x} (\tau, \xi)} \frac{1}{z^{2}} \, 
k_{i \frac{\Omega}{a}} (z) e^{- i {\bf k}_{\perp} {\bf x}_{\perp}} = \int \frac{dk_1}{\sqrt{2 \pi}} \; \frac{1}{\sqrt{2 \omega_{k}}} \,\sum^{\,d-1}_{\Lambda = 1} \tilde{e}^{\Lambda}_{1} (k_1, {\bf k}_{\perp}) \tilde{a}_{\Lambda} (k_1, {\bf k}_{\perp}) \nonumber \\
&& \cdot \Big \{
\frac{\cosh a \tau}{k_{\perp}} \int^{\infty}_{0}  \frac{dz}{z^{2}} \; k_{i \frac{\Omega}{a}} (z) 
e^{i \kappa(\tau)z}    
%\\ & &  
- \frac{a}{k_{\perp}} \int^{\tau} d \tau^{\prime} \sinh a \tau^{\prime} \int^{\infty}_{0}
\frac{dz}{z^{2}} \;  k_{i \frac{\Omega}{a}} (z) e^{i \kappa(\tau^{\prime}) z} 
 \nonumber \\
 && -i \; \frac{a}{k_{\perp}} \int^{\tau} d \tau^{\prime} \sinh a \tau^{\prime} \sinh (\beta_{\bf k}- a \tau^{\prime})
\int^{\infty}_{0} \frac{dz}{z} \;  k_{i \frac{\Omega}{a}} (z) 
e^{i\kappa (\tau^{\prime})z} \Big \}
+ (\text{h.c.},\; {\bf k}_{\perp} \to - {\bf k}_{\perp}) \,. 
\label{kap1}
\end{eqnarray}
The quantities $\kappa$ and $\beta_{\bf k}$  are defined in Eq.(\ref{beta}).
The above expression contains two divergent integrals over $z$ near $z = 0$.
However, a partial integration of the second one with respect to $\tau$ 
gives an integrated term which cancels the first one leaving a convergent 
integral over $z$. This contribution together with the other convergent term 
gives
\begin{eqnarray}
&&\int d \xi d^{\,d-1} x_{\perp} \hat{\tilde{A}}_{1}^{\prime} 
(t,{\bf x}) \; \Big |_{t,{\bf x} = t,{\bf x} (\tau, \xi)} \frac{1}{z^{2}} \, 
k_{i \frac{\Omega}{a}} (z) e^{- i {\bf k}_{\perp} {\bf x}_{\perp}}= i \int^{\infty}_{- \infty} \frac{dk_1}{\sqrt{2 \pi}} \; \frac{1}{\sqrt{2 \omega_{k}}} \, 
\; \frac{\omega_{k}}{k_{\perp}^2} \, 
\frac{a^{2}}{\Omega}  \nonumber\\
&&\cdot \sqrt{\frac{2}{a \Omega \sinh \pi \frac{\Omega}{a}}} \,   \sum^{\,d-1}_{\Lambda = 1}\Bigg\{ \sin \frac{\Omega}{a} \Big( \beta_{{\bf k}} - a \tau - \frac{i \pi}{2} \Big) 
\tilde{e}^{\Lambda}_{1} (k_1, {\bf k}_{\perp}) \tilde{a}_{\Lambda} (k_1, {\bf k}_{\perp}) \nonumber\\
&& + \sin \frac{\Omega}{a} \Big( - \beta_{{\bf k}} + a \tau - 
\frac{i \pi}{2} \Big) \tilde{e}^{\star\,\Lambda}_{1} (k_1, -{\bf k}_{\perp}) \tilde{a}^{\dagger}_{\Lambda} (k_1,- {\bf k}_{\perp}) \Bigg\} .
\label{4}
\end{eqnarray}
Identification of Eqs.\,(\ref{2})  and\,(\ref{4}) establishes the relation between the creation and annihilation operators in the  two coordinate systems, 
\begin{eqnarray*}  
a_{1} (\Omega, {\bf k}_{\perp})&=&\frac{1}{\sqrt{a \sinh \pi \frac{\Omega}{\pi}}}  \int 
\frac{dk_1}{\sqrt{2 \pi}} \; \frac{1}{\sqrt{2 \omega_{k}}}  \; \frac{\omega_{k}}{k_{\perp}} \;
e^{i \frac{\Omega}{a} \beta_{{\bf k}}}
\nonumber\\
&\cdot& 
 \sum^{\,d-1}_{\Lambda = 1}\left [ e^{\frac{\pi \Omega}{2a}}\tilde{e}^{\Lambda}_{1} (k_1, {\bf k}_{\perp}) \tilde{a}_{\Lambda} (k_1, {\bf k}_{\perp})  - e^{- \frac{\pi \Omega}{2a}}\tilde{e}^{\star\,\Lambda}_{1} (k_1, -{\bf k}_{\perp}) \tilde{a}^{\dagger}_{\Lambda} (k_1,- {\bf k}_{\perp}) 
 \right ].
\label{5}
\end{eqnarray*} 
The relation of the perpendicular  gauge field components are  obtained in a similar  manner. 
The  final result for  the relation between the creation and annihilation operators in the inertial and the accelerated frames  (cf.\,the corresponding expression (\ref{alpa}) for the scalar field) reads
\begin{equation}  
a_{i} (\Omega, {\bf k}_{\perp})  =   \frac{1}{\sqrt{a \sinh \pi \frac{\Omega}{a}}}  \int 
\frac{dk_1}{\sqrt{2 \pi}} \; \frac{1}{\sqrt{2 \omega_{k}}}  \,
e^{i \frac{\Omega}{a} \beta_{\bf k}} 
\left [ e^{\frac{\pi \Omega}{2a}} \tilde{b}_{i} (k_1, {\bf k}_{\perp}) + e^{- \frac{\pi \Omega}{2a}}
\tilde{b}^{\dagger}_{i} (k_1, - {\bf k}_{\perp}) \right ]\,.
\label{abt}
\end{equation}
The appearance of the 2 systems of polarization vectors, the $d-2$  dimensional basis of polarization vectors ${\bf e}^{L}({\bf k}_{\perp})$ in Eq.\,(\ref{pol})  and the $d-1$ dimensional basis $\tilde{\bf e}^{\Lambda}(k_1,{\bf k}_{\perp})$ in Eq.\,(\ref{aht}), made the following redefinition of the creation and annihilation operators necessary
\begin{equation}
\tilde{b}_{L} (k_{1}, {\bf k}_{\perp}) = \sum^{\,d-1}_{I = 2} \sum^{\,d-1}_{\Lambda = 1} 
\tilde{e}^{\Lambda}_{I} (k_{1},{\bf k}_{\perp}) e^{L}_{I} ({\bf k}_{\perp}) \tilde{a}_{\Lambda} 
(k_{1}, {\bf k}_{\perp})\, ,\quad 2 \le L \le d - 1\, ,
\label{bi}
\end{equation}
and the  rescaling 
\begin{equation}
\tilde{b}_{1} (k_{1}, {\bf k}_{\perp}) = \frac{\omega_k}{k_{\perp}} \; \sum_{\Lambda=1}^{\,d-1}\tilde{e}_1^{\Lambda} (k_1,{\bf k}_{\perp})\, \tilde{a}_{\Lambda}(k_1,{\bf k}_{\perp}) 
 \, ,
\label{b1}
\end{equation}
 resulting in a system of creation and annihilation operators with standard commutation relations
\begin{equation}
 \big [ \tilde{b}_{i} (k_{1}, {\bf k}_{\perp}) , \; \tilde{b}^{\dagger}_{j} 
(k^{\prime}_{1}, {\bf k}^{\prime}_{\perp}) \big ]=\delta_{ij}\delta (k_{1} - k^{\prime}_{1}) \delta ({\bf k}_{\perp} - {\bf k}^{\prime}_{\perp}) \, \quad i, j =1,\ldots,d-1\, .
\label{bco}
\end{equation}  
As a consequence, the result (\ref{urt2}) for the scalar field applies directly to the Maxwell field and we find for the average particle number for the $d-1$ degrees of freedom  in the inertial frame ground state 
\begin{equation}
\langle 0_M| a^{\dagger}_{i} (\Omega, {\bf k}_{\perp}) a_{j} (\Omega^{\prime}, {\bf k}_{\perp}^{\prime}) | 0_M \rangle
=\frac{1}{e^{2 \pi \frac{\Omega}{a}} - 1} \delta_{i j}\,\delta (\Omega - \Omega^{\prime})\delta({\bf k}_{\perp}-{\bf k}^{\prime}_{\perp}) \, .        
\label{urtmax}
\end{equation}
With the relation (\ref{abt}) between the Rindler and Minkowski space creation and annihilation operators,  we can proceed as in the case of the scalar field \cite{UNRU76},\cite{UNWA84}. The relation of the left Rindler wedge operators with the Minkowski operators differs from the relation  (\ref{abt}) only in the sign of the phase $\Omega\beta_{\bf k}/a$. Given these two sets of equations the Minkowski vacuum can be expressed in terms of correlated  product states in $R_+$ and $R_-$ 
\begin{equation}
  \label{unwa}
 |0_M\rangle = \prod_{\alpha} {\cal N}_{\alpha} \sum_{n_{\alpha}} 
e^{-\pi n_{\alpha}\Omega_{\alpha}/a }|n_{\alpha}\rangle _+ \otimes |n_{\alpha}\rangle _- \,,\quad {\cal N}_{\alpha} = \sqrt{1-e^{-2\pi\Omega_{\alpha}/a}} \,.
\end{equation}
The structure of this relation is the same as for the scalar field \cite{UNWA84}. The index $\alpha$ specifies besides transverse momenta ${\bf k}_{\perp}$ and energy $\Omega$ (cf.\,Eq.\,(\ref{abt})) the $d-1$ polarization states.  For observables built exclusively from degrees of  freedom in $R_+$,  the matrix elements are given in terms of the density matrix (\ref{denma}) with the Hamiltonian (\ref{Eng2})  of the Maxwell field. \\
We illustrate this result by computing the energy density of the Maxwell field. 
As for thermal expectation values we calculate the expectation values in the Minkowski vacuum  by assuming the operators to be normal ordered with respect to the Rindler vacuum. Using the normal mode expansion of the gauge fields (\ref{notr1a}), (\ref{trnoi}) the properties of the polarization vectors (\ref{pol}) and the above relation between Rindler and Minkowski space creation and annihilation operators we find for the electric and magnetic energy densities (cf. Eq.\,(\ref{hapr3})) 
\begin{eqnarray}
\langle 0_M | {\cal H}_E |0_M\rangle &=&\frac{1}{2}\langle 0_M |:\Big\{ e^{2a\xi} \hat{\Pi}^{1\,2}+ \sum_{I=2}^d \hat{\Pi}^{I\,2}\Big\}:|0_M\rangle\nonumber\\ &=& (d-1) a^{d-2} e^{(1-d)a\xi} \int d\omega \frac{1}{e^{2\pi\frac{\omega}{a}}-1} \Big[\omega^2 +\frac{a^2}{2} (d-1) \Big] I(d-2,\omega) \nonumber
\\ \langle 0_M | {\cal H}_B |0_M\rangle &=&\frac{1}{2}\langle 0_M |:\sum_{I=2}^d \Big\{ F_{1I}^2 + e^{2a\xi}\sum_{J>I}^d  F_{IJ}^2 \Big\}:|0_M\rangle\nonumber\\ &=& (d-1) a^{d-2} e^{(1-d)a\xi} \int d\omega \frac{1}{e^{2\pi\frac{\omega}{a}}-1} \Big[\omega^2 +\frac{a^2}{2} (d-1)(d-2) \Big] I(d-2,\omega)\,,\nonumber\\ 
\label{hEB}
\end{eqnarray}
with 
 (cf. \cite{RG65})
\begin{displaymath}
I (\lambda, \omega) = \frac{a\Omega_{\lambda}}{2\omega (2 \pi)^{\lambda+1}} \int ^{\infty}_{0} x^{\lambda} 
k^{2}_{i\frac{\omega}{a}} (x) \, d x 
= \frac{\Omega_{\lambda}}{(2 \pi)^{\lambda+3}} \,\sinh \pi \;\frac{\omega}{a}\, 2^{\lambda}
\frac{\Gamma^{2} (\frac{1+\lambda}{2})}{\Gamma (1+\lambda)} \; \Big | \Gamma  \Big ( 
\frac{1+\lambda}{2} + i \, \frac{\omega}{a} \Big ) \Big | ^{2} \,,
\end{displaymath}
and $\Omega_{\lambda}$ denotes the volume of the $\lambda$-dimensional sphere. For $d=3$ our result for the energy density 
\begin{equation}
\langle 0_M | {\cal H}_E +  {\cal H}_B |0_M\rangle = \frac{1}{\pi^2}   e^{-2a\xi} \int \omega d\omega \frac{\omega^2 +a^2}{e^{2\pi\frac{\omega}{a}}-1}
\label{hEB3}
\end{equation}
agrees  with  the results obtained in \cite{CADE77}, \cite{SAHA02}\, and with the result for classical random radiation in \cite{BOYE80}. The spectrum differs from the Planck spectrum, i.e. the density of states is not that of black-body radiation.  In agreement with earlier findings \cite{TAGA86}  we  note that  as a result of the integration over  ${\bf k}_{\perp}$  (cf.\,\cite {UNRU86}) 
and the property
\begin{equation}
  \label{sico}
\frac{ \Big | \Gamma  \big ( 
\frac{1}{2} + i \mu \big ) \Big| ^{2} }{ \Big | \Gamma  \big ( 
1 + i \, \mu\Big ) \Big | ^{2} }\; \frac{1}{e^{2\pi\mu}-1} = \mu^{-1} \frac{1}{e^{2\pi\mu}+1}\,,
\end{equation}
the characteristic Bose-Einstein denominator is replaced in odd space-time dimensions $d+1$ by the Fermi-Dirac denominator $1+\exp(2\pi\omega/a)$\,.\\
From Eq.\,(\ref{hEB}) we conclude that the action density 
\begin{equation}
\langle 0_M | s |0_M\rangle = \langle 0_M | {\cal H}_E -  {\cal H}_B |0_M\rangle = \frac{(d-1)(3-d)}{4} a^{d} e^{(1-d)a\xi} \int d\omega \frac{1}{e^{2\pi\frac{\omega}{a}}-1}  I(d-2,\omega)
\label{acd}
\end{equation}
vanishes for $d=3$,  is positive for $d=2$ and negative for $d\ge 4$. Thus in Rindler space, the Maxwell field develops in general a non-vanishing photon condensate.  In 5 and higher dimensional Rindler spaces the photon condensate is dominated by the magnetic fluctuations as is the case for the  gluon condensate in (4-dimensional) QCD \cite{SVZ79}. 
\subsection{Gauge fields coupled to external currents} \label{sec:riex}
Here we discuss the relation between the gauge fields in inertial and accelerated frames if external currents are present. This is an important  topic when comparing the observations of inertial and accelerated observers. In particular the absence of bremsstrahlung  for the  co-accelerated observer has been a central  issue in this discussion. \\
 As above we start with the identification  of the Heisenberg operators in the Rindler space and in the right Rindler wedge $R_+$ of Minkowski space.  According to the results of Secs.\,\ref{sec:decpl} and \ref{sec:tiev},  the  identification of the fields involves on the one hand the Heisenberg operators of the Maxwell field in the absence of sources and on the other  the (c-number) solutions of the classical Maxwell equations coupled to the external currents. Therefore the identification (\ref{aphtph2}) of the transverse gauge field operators remains valid together with the relation (\ref{abt}) between creation and annihilation operator. The c-number part is obtained by  solving  the  equations of motion (\ref{eqri}) for the classical transverse gauge field $\hat{\alpha}^i$ which in Rindler space read (cf.Eqs.\,(\ref{notr1}),\,(\ref{notri}))
\begin{eqnarray*}
\partial_{\tau}^{2}\hat{\alpha}^{1} -\Delta_s \hat{\alpha}^{1} &=&e^{2a\xi}j^1 +\partial^{1} \frac{1}{\Delta} \partial_{j} (e^{2a\xi} j^{j})\, ,\\
%\label{notr2a}
\partial_{\tau}^{2}\hat{\alpha}^{I} -\Delta_s \hat{\alpha}^{I} -2ae^{2a\xi}\partial^I\hat{\alpha}^{1}&=&e^{2a\xi}j^I +\partial^{I} \frac{1}{\Delta} \partial_{j} (e^{2a\xi} j^{j})\, .
\label{notria}
\end{eqnarray*} 
The c-number solutions of the Maxwell equation in the two frames are easily identified. We may for instance solve the equations of motion in Minkowski space, perform the coordinate transformation to Rindler space, carry out the necessary gauge transformations to the Weyl gauge and compute the transverse gauge field and the electric field.  In this procedure it is essential that in the space of physical states, the eigenvalue of the longitudinal electric field operator is a solution of  the Maxwell equations. Furthermore the  gauge freedom of the auxiliary fields $\alpha$ (Eq.\,(\ref{eqch0})) and the freedom in the choice of the time independent integration constants $E^i_{(\ell)}({\bf x})$ in Eq.\,(\ref{GLC}) and $\chi_0(\omega,{\bf k}_{\perp})$ to be derived below (Eq.\,(\ref{chj})) are essential for establishing the relation between the classical fields in Minkowski and Rindler spaces. \\
Given the general structure (\ref{heis}) of the  Heisenberg field operators after  decoupling of the  external currents and given the relation (\ref{abt}) between Minkowski and Rindler space  creation and annihilation operators, it is obvious that at the level of fields or bilinears of  fields the classical considerations concerning the relation between Minkowski and Rindler space observables are essentially unchanged. In the case of the energy momentum tensor the classical result is modified only by an additive contribution from the Unruh radiation of the free Maxwell field.   In particular, the resolution of the equivalence principle paradox concerning the absence of radiation in a co-accelerated frame given in \cite{BOUL80} remains valid (cf.\,also \cite{HIMA93}).

The interplay between classical and quantum fields becomes more subtle in  the time evolution of the states. For this purpose it is convenient to express the unitary transformation $W$ (\ref{shif}) in terms of the amplitudes of the  normal modes introduced in Sec.\,\ref{sec:trgf}.  We start with the identity (\ref{dpht})  (cf.\,also Eq.\,(\ref{trcu}))
\begin{equation}
\frac{d}{d\tau}\hat{\Phi}_{H} 
  = - \int  d\xi e^{2a\xi} d^{\,d-1} x_{\perp} \hat{A}_{H i} j^i\,.
\end{equation}
Written  in terms of  the normal mode expansions (\ref{notr1a}, \ref{trnoi}) this equation reads
\begin{displaymath}
\frac{d \hat{\Phi}_{H}}{d\tau} = - \int d \omega \int d^{\,d-1} k_{\perp} \sum^{\,d-1}_{i = 1} 
\Big ( a_{i} (\omega, {\bf k}_{\perp}) e^{- i \omega \tau} \hat{J}^{i} (\tau, \omega, {\bf k}_{\perp}) 
+ {\rm h.~c.} \Big ) \,,
\end{displaymath}
with
\begin{eqnarray}
\hspace{-.5cm}\hat{J}^1 (\tau, \omega, {\bf k}_{\perp})&=&\frac{1}{\sqrt{2 \omega^{3}}}\,\int \frac{d^{\,d-1}x_{\perp}}{(2 \pi)^{(d-1)/2}} 
e^{i {\bf k}_{\perp} {\bf x}_{\perp}} \int^{\infty}_{0} dz \, \frac{a^2 z^2}{k_{\perp}^2}  
\nonumber\\ 
&&\cdot \Big ( \frac{a}{k_{\perp}} z j^{1} (\tau, z, {\bf x}_{\perp}) +i\frac{k_{I}}{k_{\perp}} j^{I} (\tau, z, {\bf x}_{\perp})\ \frac{d}{dz} \Big ) \, k_{i\frac{\omega}{a}} (z)\,,
\label{jr1}\\
\hspace{-.5cm}\hat{J}^{L} (\tau, \omega, {\bf k}_{\perp})&=&\frac{1}{\sqrt{2 \omega}} \int \frac{d^{\,d-1}x_{\perp}}{(2 \pi)^{(d-1)/2}} \, 
e^{i {\bf k}_{\perp} {\bf x}_{\perp}} \int dz \frac{a z}{k_{\perp}^2}\,   %
\sum^{d}_{I = 2} e^{L}_{I} ({\bf k}_{\perp}) j^{I} (\tau, z, {\bf x}_{\perp})k_{i\frac{\omega}{a}} (z) \,.\nonumber\\\label{jrL}
\end{eqnarray}
Integration over time 
and  transformation  back to Schr\"odinger operators yields  
\begin{displaymath}
\hat{\Phi} (\tau) = - \int d \omega \int d^{\,d-1} k_{\perp} \sum^{\,d-1}_{i = 1} \Big (a_{i} (\omega, {\bf k}_{\perp})
\chi^{ i} (\tau, \omega, {\bf k}_{\perp}) + {\rm h.~c.} \Big )\,,
\end{displaymath}
with
\begin{equation}
\chi^{i} (\tau, \omega, {\bf k}_{\perp}) = \int^{\tau} d \tau^{\prime} \, e^{ i \omega (\tau - \tau^{\prime})}
\hat{J}^{i} (\tau^{\prime}, \omega,{\bf k}_{\perp})+\chi^{i}_0 (\omega, {\bf k}_{\perp}) e^{i\omega\tau}\,.
\label{chj}
\end{equation}
The arbitrariness in the time integration of  $\hat{\Phi}_{H} $ reflects the ambiguity  in the choice of Green's function. The  choice $\chi^{i}_0 (\omega, {\bf k}_{\perp})=0$ corresponds to the retarded and 
$$\chi^{i}_0 (\omega, {\bf k}_{\perp})= -\int^{\infty} d \tau^{\prime} \, e^{ -i \omega \tau^{\prime}}
\hat{J}^{i} (\tau^{\prime}, \omega,{\bf k}_{\perp})$$ 
to the advanced propagator.  \\ 
 The time evolution in the longitudinal and in the 
transverse sector of the Hilbert space are independent of each other. The unitary  transformation $W$ (\ref{shif}) is 
a product of operators acting separately on  the state vectors in the corresponding sector  
\begin{displaymath}
W(\tau) = e^{i \hat{\Phi}(\tau)} e^{i \Phi_{(\ell)} (\tau)}\,,\quad 
| \psi (\tau) \rangle = | \hat{\psi} (\tau) \rangle\, | \psi _{(\ell)} (\tau) \rangle \,, 
\end{displaymath}
with  $| \psi _{(\ell)}\rangle$ being eigenstates of the longitudinal electric field.     
Since   $| \psi_{(\ell)} (\tau_{0}) \rangle $   is a state with vanishing electric field
we conclude that at time $\tau$
\begin{displaymath}
\Pi^i_{(\ell)}| \psi_{(\ell)} (\tau) \rangle =  - \partial^{i} \frac{1}{\Delta} \; e^{2 a \xi} j^{0} (\tau,\xi, {\bf x}_{\perp}) | \psi_{(\ell)} (\tau) \rangle\,, 
\end{displaymath}
i.e. the longitudinal electric field is given by the Coulomb field generated  
(instantaneously) by the charge density $j^{0}$. \\
With  the ground state as the initial state we find 
 \begin{equation}
e^{i \hat{\Phi}(\tau)} e^{- i H_{0} (\tau-\tau_{0})} | 0 \rangle = C(\chi) \exp\Big\{- i \int d \omega\, d^{\,d-1} k_{\perp} \sum^{\,d-1}_{i = 1} a^{\dagger}_{i} (\omega, {\bf k}_{\perp}) 
\chi^{i\,\star} (\tau, \omega, k_{\perp}) \Big\} | 0 \rangle \,,
\label{foco}
\end{equation}
where $C(\chi)$ arises from normal ordering $\exp \,\hat{\Phi} $
\begin{equation}
C(\chi)=\exp\Big\{- \frac{1}{2} \int d \omega d^{\,d-1} k_{\perp} |\chi^{i} 
(\tau, \omega, {\bf k}_{\perp})|^2 \Big\} \,.
\label{cnoo}
\end{equation}
By projection  of the time evolved ground state on the relevant Fock components,  the associated probabilities and henceforth the  photon production rates are easily calculable without having to invoke perturbation theory in the coupling. 

Formally the above expression together with the definition   (\ref{chj}) of $\chi$ apply to the time evolution in Minkowski space as well,  with the  Fourier components  given by
\begin{equation}
  \label{jmi}
  \hat{J}^{L} (t,{\bf k}) = \int \frac{d^{\,d}x}{(2 \pi)^{d/2} \sqrt{2 \omega_{\bf k}}} \, 
e^{i {\bf k} {\bf x}}  
\sum^{d}_{I = 1} e^{L}_{I} ({\bf k}) \hat{j}^{I} (t,{\bf x})\,,\quad L=1,\ldots, d-1.
\end{equation}
The study of the time evolution is pertinent to the  issue  of  bremsstrahlung generated by a uniformly accelerated charge and observed either in the inertial or in the co-accelerated frame.  
The probability to observe the system at the time $\tau $ in a state characterized by the occupation numbers
$m_{\alpha}$, with $\alpha$ standing for $\omega$, ${\bf k}_{\perp}$ and the polarization,  is given by  (cf.\,Eqs.\,(\ref{denma}),\,(\ref{unwa}))
\begin{equation}
P (0_{M} \to \{ m_{\alpha}\} ) = {\rm tr}\Big[ \prod_{\alpha} | m_{\alpha} \rangle \langle m_{\alpha}| \rho (\tau)\Big]
= \prod_{\alpha} {\cal N}_{\alpha}^2 \sum_{n_{\alpha}} e^{- 2 \pi n_{\alpha} \Omega_{\alpha}/a}
|M_{m_{\alpha}, n_{\alpha}}|^2 \,.                                 
\label{pmn}
\end{equation}
After a straightforward calculation one obtains the following closed expression for these  
matrix elements 
\begin{equation}
M_{m_{\alpha}, n_{\alpha}} = (-i)^{n_{\alpha}+m_{\alpha}}e^{- \frac{1}{2} |\chi_{\alpha}|^{2}}  \sum^{\mu (n_{\alpha}, m_{\alpha})}_{k = 0}
\frac{\sqrt{n_{\alpha}! \,m_{\alpha}!}}{(n_{\alpha} - k)! (m_{\alpha}-k)!} \, \frac{(-1)^{k}}{k!} \; 
\chi^{m_{\alpha} - k}_{\alpha} \chi^{\star n_{\alpha}-k}_{\alpha}\,,            
\label{Mmn}
\end{equation} 
with
\begin{displaymath}
\mu (n_{\alpha}, m_{\alpha}) =  \mbox{Min}\,(n_{\alpha}, m_{\alpha}) \,.
\end{displaymath}
Expression (\ref{pmn}) together with the matrix elements (\ref{Mmn}) contain the complete  information about
the time evolution.  
In the perturbative limit only the term $k = \mu (n_{\alpha}, m_{\alpha})$ is retained and the one photon emission  and absorption probabilities become
\begin{equation}
  \label{1pht}
|M_{n_{\alpha} + 1, n_{\alpha}}|^2 \approx (n_{\alpha}+1)\,|\chi_\alpha|^2 \,,\quad |M_{n_{\alpha} - 1, n_{\alpha}}|^2 \approx n_{\alpha} |\chi_\alpha|^2\,.   
\end{equation}

Most of the discussion on the connection between Rindler and Minkowski space have been carried out in the context of scalar field theory.  We conclude this section by compiling the equations relevant for such studies.
According to Eq.\,(\ref{unsc2}), the exponent of the unitary transformation $w$ (\ref{unsc}) satisfies in Rindler space the differential equation  
\begin{displaymath}
\frac{d \varphi_{H}}{d\tau} = -\int e^{2a\xi} d \xi \,d x_{\perp} \rho \phi   = -\int d \omega d^{d-1} k_{\perp} \Big (a (\omega, k_{\perp}) \, 
e^{- i \omega \tau} \rho (\tau, \omega, {\bf k}_{\perp}) + {\rm h.c.}  \Big )\,,
\end{displaymath}
where we have used the normal mode expansion in Eq.\,(\ref{nomosc}) and have introduced  
\begin{equation}
\rho (\tau, \omega, {\bf k}_{\perp}) = \frac{1}{\sqrt{2 \omega}\,k_{\perp}^2} \int az dz  \; 
\frac{d^{d-1}x_{\perp}}{(2 \pi)^{(d-1)/2}} \; e^{i {\bf k}_{\perp} {\bf x}_{\perp}} 
k_{i\frac{\omega}{a}} (z)\rho(\tau,z,{\bf x}_{\perp}) \,.
\label{rhosc}
\end{equation}
Carrying out the time integration and transforming back to Schr\"odinger operators we find
\begin{displaymath}
\varphi (\tau) = \int d \omega \, d^{d-1} k_{\perp} \Big ( a (\omega, {\bf k}_{\perp}) \chi 
(\tau, \omega, k_{\perp}) + {\rm h.c.} \Big ) \,,
\end{displaymath}
\begin{equation}
\chi (\tau, \omega, {\bf k}_{\perp}) = \int^{\tau} d \tau^{\prime} \, e^{i \omega (\tau - \tau^{\prime})}
\rho (\tau^{\prime}, \omega, k_{\perp})+\chi_0 (\omega, {\bf k}_{\perp})e^{i\omega\tau} \,.
\label{chisc}
\end{equation}
The c-number contribution to the energy (\ref{hxs2})
is given by
\begin{equation}
  \label{hscri}
h^{sc}(\tau) = \frac{1}{2} \int e^{2a\xi}\ d\xi\, d^{d-1}x_{\perp}\rho(\tau,\xi,x_{\perp}) \alpha(\tau,\xi,x_{\perp})\,,   
\end{equation}
with the solution $\alpha$ of the inhomogeneous wave equation (cf.\,Eqs.\,(\ref{EQM1}) and (\ref{scala}))
\begin{equation}
\Big [ \partial^{2}_{\tau} - \Delta_s + m^{2} \;
e^{2 a \xi} \Big ]  \; \alpha = - e^{2a\xi}\rho   \, .
\label{EQMin}
\end{equation}
\subsection{Initial value problems and bremsstrahlung}\label{ivp}
Important for understanding the relation between physics in Rindler and Minkowski spaces is the issue of  bremsstrahlung of a uniformly accelerated charge. While observed in an inertial frame, a co-accelerated observer will measure a static electric field rather than photons. In the context of classical physics \cite{BOUL80} this puzzle  has been resolved by the observation that the radiation is emitted into a region which is inaccessible to the co-accelerated observer. In a quantum mechanical treatment and considering the static charge as a limit of an accelerated charge, some remnant of the radiation has been found to persist  \cite{HIMS921,HIMS922} and to get enhanced by the divergence of the occupation probability  (cf.\,Eq.\,(\ref{urtmax})) in the infrared limit. The net result is an emission and absorption  rate of photons in the ``co-accelerated '' frame whose sum is identical to the production rate in the inertial frame. This method has been applied also to the case of a uniformly accelerated scalar source \cite{MATS96}\,, \cite{REWE94} and has been extended in a calculation of the emission rate of a Unruh-DeWitt detector \cite{DIST02}. We will verify these results within our approach and in order to avoid difficulties associated with an ``eternal'' acceleration we  will  analyze the relation between the physics in Rindler and Minkowski spaces in the presence of a uniformly accelerated charge by formulating an appropriate initial value problem. 
\subsubsection{Scalar radiation}
We assume a pointlike scalar source to be located at the  fixed position $\xi=0,\, {\bf x}_{\perp}=0$ (cf.\,Eq.\,(\ref{rhosc})) 
\begin{equation}
\rho (\tau, \omega, {\bf k}_{\perp}) =q_0 \sigma(\tau)\rho (\omega, {\bf k}_{\perp}) = \frac{q(\tau)}{\sqrt{2 \omega}} \; k_{i\frac{\omega}{a}}
(k_{\perp}/a) \; \frac{1}{(2 \pi)^{(d-1)/2}} \,,
\label{rhosc2}
\end{equation}
with its  strength 
$$q(\tau) = q_0 \sigma(\tau)$$  
being turned on in the interval
\begin{equation}
\tau_0\le \tau \le \tau_0+\delta\tau\,,
\label{adint} 
\end{equation}
with
$$ \sigma(\tau)=0\; \text{for}\; \tau\le \tau_0\,,\quad \sigma(\tau)\approx 1 \; \text{for}\; \tau\ge \tau_0+\delta \tau\,. $$
For times later than $\tau_0+\delta \tau$,  $\sigma(\tau)$ may differ from the asymptotic value by exponentially suppressed corrections as is the case with the following ansatz
$$\sigma(\tau)=\theta(\tau-\tau_0)\,(1-e^{-\lambda (\tau-\tau_0)})\,,\quad \lambda \sim 1/\delta\tau\,.$$
In perturbation theory, the  rate of emission of one particle with frequency $\omega_{\alpha}$ is given by (cf.\,Eq.\,(\ref{1pht}))
\begin{equation}
 r_+(\omega_{\alpha},{\bf k}_{\perp})=(n_{\alpha}+1)\,\frac{d|\chi_\alpha|^2}{d\tau}=  (n_{\alpha}+1)\,q_{0}^2\,\rho (\omega_{\alpha}, {\bf k}_{\perp})^2\,2\sigma(\tau)\int_{\tau_0}^{\tau}d\tau^{\prime} \frac{\sin \omega_{\alpha}(\tau-\tau^{\prime})}{\omega_\alpha}\frac{d\sigma(\tau^{\prime})}{d\tau^{\prime}} \,.
\label{emra1}
\end{equation} 
We have used $\chi_0=0$ and carried out an integration by parts. In this way the integral effectively extends only over the time interval (\ref{adint}). At asymptotic times $\tau\gg \tau_0+\delta\tau$ only frequencies $\omega_{\alpha}\ll 1/\delta\tau$ contribute, i.e.  turning on  the source affects only long wavelength modes and we can write
\begin{equation}
 r_+(\omega_{\alpha},{\bf k}_{\perp})\approx 2 (n_{\alpha}+1)\,q_{0}^2\,\rho (\omega_{\alpha}, {\bf k}_{\perp})^2 \frac{\sin \omega_{\alpha}(\tau-\tau_0)}{\omega_\alpha}\rightarrow  2\pi (n_{\alpha}+1)q_0^2\rho (\omega_{\alpha}, {\bf k}_{\perp})^2 \delta(\omega_{\alpha}) \,.
\label{emra2}
\end{equation}
At this point we essentially can follow the arguments given in \cite{REWE94}. We note that for $\omega \ll a$ (cf.\,Eq.\,(\ref{kmu}),\,(\ref{rhosc}))
\begin{equation}\rho(\omega,{\bf k}_{\perp}) \rightarrow \frac{\sqrt{2\omega}}{a\,(2\pi)^{d/2}}K_{0}(k_{\perp}/a)\,,
\label{ro0}
\end{equation}
and therefore the rate $r(\omega_{\alpha},{\bf k}_{\perp})$ vanishes. However  the divergent occupation probability in the Unruh heat bath compensates the infrared decrease of the rate. The total rate observed long after the source has been switched on is given by
\begin{equation}
 r_+({\bf k}_{\perp})= \int_0^{\infty} d\omega (1-e^{-2\pi\omega/a})\sum_{n_{\omega}=0}^{\infty}(n_{\omega}+1) e^{-2\pi n_{\omega}\omega/a} r_+(\omega,{\bf k}_{\perp})= \frac{q_0^2}{a (2\pi)^d} K_0^2(k_{\perp}/a)\,.
\label{tora}
\end{equation}
Due to the infrared enhancement by the Unruh heat bath, the emission rate remains affected at arbitrary late times by the process of turning on the source.  
The absorption rate $r_-$ (cf.\,Eq.\,(\ref{1pht})) is identical to the emission rate and, in agreement with the result of \cite{REWE94}  the total rate $r=r_++r_-$ can be shown to coincide with the one particle production rate. This is a straightforward calculation in Minkowski space with the scalar source given by (here the corrections from switching on the source are negligible for sufficiently late times)
\begin{displaymath}
\rho (t, {\bf k}) = \frac{1}{(2 \pi)^{d/2}\sqrt{2 \omega_{{\bf k}}}} \; q \, e^{i k_{1} \sqrt{t^{2} + 1/a^{2}}}
\frac{1}{a \sqrt{t^{2} + 1/a^{2}}} \,.
\end{displaymath} 
\subsubsection{Photons}
In Weyl gauge, the source of the radiation is the spatial part of the current. If generated  by a point charge at rest at $\xi = 0$ in the accelerated frame the only non vanishing spatial component of the current in the inertial frame is given by  
\begin{equation}
 j^1 (t,x,{\bf x}_{\perp}) = q_0 \frac{a^2 t}{\sqrt{1 + a^2 t^2}} \delta({\bf x}_{\perp})
\delta\big(ax-\sqrt{a^2t^2 + 1}\big)\,.
\end{equation}
From Eqs.\,(\ref{jmi}) and (\ref{chj}) it is straightforward to derive the following expression for $\chi^1$ 
\begin{equation}
 \big|\chi^{1} (t, {\bf k})\big|^2 = \frac{q_0^2}{(2\pi)^{d} 2 k} 
\frac{{k_{\perp}}^2}{k^2} 
\Big|\int_{} ^{t} dt' \frac{a t'}{\sqrt{1 + a^2 {t'}^2}} 
e^{-ikt' + ik_1 \sqrt{{t'}^2 + \frac{1}{a^2}}}\,\Big|^2\,.
\end{equation}
From here on we follow the computation  \cite{HIMS922}  of 
 the total emission rate of photons with given transverse momentum ${\bf k}_{\perp}$.   By integrating the above expression over the 1-component of the photon momenta and dividing by the proper time during which the charge is accelerated we obtain
\begin{equation}
r(k_{\perp}) = \frac{1}{T(t)}\int_{-\infty}^{\infty}dk_1\big|\chi^{1} (t,k_1, {\bf k}_{\perp})\big|^2 = \frac{2 q_0^2}{(2\pi)^{d} a} \Big|K_1 \Big(\frac{k_{\perp}}{a}\Big)\Big|^2\,.
\label{rmin}
\end{equation}
Changes in $r(k_{\perp})$ due to switching on or off the current are negligible if the acceleration lasts for a sufficiently long (proper) time $T(t)$.

In electrodynamics, current conservation requires a more complicated procedure for comparing photon emission and absorption rates in accelerated and inertial frames. We adopt the dipole arrangement of \cite{HIMS921,HIMS922} with the following current in the accelerated frame 
\begin{eqnarray}
j^0 = q_0 \sigma (\tau) e^{-2a\xi} [\delta (\xi) - \delta (\xi - L)] 
\delta({\bf x}_{\perp})\,,\quad 
j^1 = q_0 \dot{\sigma} (\tau) e^{-2a\xi} \theta (\xi) \theta (L - \xi) 
\delta({\bf x}_{\perp})\,,
\label{hipa}
\end{eqnarray}
with the time dependence $\sigma(\tau)$ to be specified later. 
In the limit  $L\to \infty$,  the normal mode component of the current defined in Eq.\,(\ref{jr1}) is  given by 
\begin{equation}
\hat{J}^1(\tau, \omega, {\bf k}_{\perp}) = 
\frac{2q_0\dot{\sigma}(\tau)}{\omega (2\pi)^{\frac{d+1}{2}}}
\sqrt{\frac{1}{a} \sinh {\frac{\pi \omega}{a}}} M(k_{\perp},\omega)\,,
\end{equation}
with
\begin{equation}
 M(k_{\perp},\omega)= -K'_{i\frac{\omega}{a}} (\frac{k_{\perp}}{a}) + 
\frac{\omega^2}{a k_{\perp}} 
\int_{\frac{k_{\perp}}{a}} ^{\infty} \frac{dz}{z} K_{i \frac{\omega}{a}}(z)\, ,
\end{equation}
which implies (cf.\,Eq.\,(\ref{chj}) with $\chi_0=0$)  the following expression for $\chi^1$  
\begin{equation}
\chi^1(\tau, \omega, k_{\perp}) 
= f(\tau,\omega) \frac{2q_0 e^{i\omega\tau}}{\omega (2\pi)^{\frac{d+1}{2}}}
\sqrt{\frac{1}{a} \sinh {\frac{\pi \omega}{a}}}\, M(k_{\perp},\omega)\,,
\label{chi1}
\end{equation}
where $f(\tau,\omega)$ is defined as
\begin{equation}
f(\tau,\omega) = \int ^{\tau} d\tau^{\prime} e^{-i\omega \tau^{\prime}}\dot{\sigma}(\tau^{\prime})\,. 
\label{fto}
\end{equation}
After including the occupation probabilities of the Rindler photons in the Minkowski vacuum, the photon emission rate (cf. Eq.\,(\ref{rmin})) becomes 
\begin{equation}
r_{+}(k_{\perp}) = \frac{1}{\tau}\int_{0} ^{\infty} d\omega 
|\chi^1 (\tau, \omega, k_{\perp})|^2
\Big(1 + \frac{1}{1-e^{-\frac{2\pi \omega}{a}}}\Big) \,.
\label{r+r}
\end{equation}
It is now easy to verify that the choice of the time dependence
\begin{equation}
\sigma(\tau)= \sqrt{2} \cos E\tau \,\theta(\tau)\theta(T-\tau)\,,
\label{hicu}
\end{equation}
and taking the consecutive limits   $T \to \infty$ and $E\to 0$   reproduces   the result obtained in \cite{HIMS921,HIMS922}
\begin{equation}
r_{+}(k_{\perp}) = \frac{q_0^2}{(2\pi)^{d} a} \big|K_1 (\frac{k_{\perp}}{a})\big|^2\,,
\label{hrp}
\end{equation}
which, after multiplication with a factor of 2  accounting  for the process of absorption of photons, agrees with the  inertial frame result (\ref{rmin}).

For interpretation of this correspondence between the results for the total rates in the inertial and accelerated frame  as an effect of switching on and off the coupling to the radiation field the modulation of the current is not necessary. This can be verified with the following choice  for $\sigma(\tau)$ 
\begin{equation}
\sigma(\tau)= \Big[1-e^{-\lambda(\tau-\tau_0)^2}\Big]\Big[1-e^{-\lambda(\tau+\tau_0)^2}\Big]\theta(\tau+\tau_0)\theta(\tau_0-\tau)\,.
\label{turn}
\end{equation}
This parametrization guarantees that no discontinuities in time and space component of the current occur when  turning on or off the coupling. Furthermore we require 
$$\lambda\tau_0^2 \gg 1\,,$$ 
such that in a time  interval of size $ 2\tau_0$ and  up to exponentially suppressed corrections ($\sim e^{-4\lambda\tau_0^2}$),   the time component of the current is constant and non-vanishing while the spatial component vanishes. Up to such exponentially suppressed contributions, we obtain for $f(\tau,\omega)$ (cf. Eq.\,(\ref{fto}) and \cite{RG65})
$$f(\tau,\omega)\approx   2\lambda \Big[ i \cos \omega\tau_0\, \frac{d}{d\omega}\sqrt{\frac{\pi}{\lambda}} e^{-\omega^2/4\lambda}+ \sin \omega\tau_0\, \frac{d}{d\omega}\Big(\frac{\omega}{\lambda}\, _1F_1\big(1;\frac{3}{2};-\frac{\omega^2}{4\lambda}\big)\Big)\Big] \,.  $$
Assuming  $\lambda$  to be  chosen sufficiently large in comparison to  the relevant frequencies this expression simplifies 
\begin{equation}
\lambda \gg \omega^2\,, \quad |f(\tau,\omega)|^2 \approx 4 \sin^2 \omega \tau_0\,,\label{af2}
\end{equation}
and assuming furthermore the  total time $2 \tau_0$ of non-vanishing $\sigma(\tau)$  to be  large on the scale of the acceleration $a\tau_0\gg 1$, the $\omega$ integration  in (\ref{r+r})  can be carried out in closed form and yields a result identical to  (\ref{hrp}). In agreement  with  \cite{HIMS921,HIMS922},  the time averaged total rates for the  parametrizations (\ref{hicu}) and (\ref{turn}) coincide with the time averaged emission rate (\ref{rmin}) of a uniformly accelerated charge in Minkowski space  as expected on the basis of the   invariance  of the  coupling term in the action $\int d^dx \sqrt{|g|}j^{\mu}A_{\mu}$ under general coordinate transformations. 

In addition to the time averaged rate we can determine the time evolution of emission and absorption probabilities  under the influence of a current whose time component is turned on at the time $-\tau_0$ and turned off at $\tau_0$. Furthermore we assume the time ($1/\sqrt{\lambda}$) needed for turning on or off to be much smaller than the total time  where the charge is turned on $(\sigma(\tau)\approx 1)$.   Since ultimately we are only interested in small frequencies (cf.\,Eq.\,(\ref{af2}))  we do not need to further specify the details of the time dependent coupling $q_0 \sigma(\tau)$ (Eq.\,(\ref{hipa})).  For  
$\omega^2\ll\lambda $,  Eq.\,(\ref{fto}) yields
\begin{equation} f(\tau,\omega) \approx \theta(\tau+\tau_0)\Big[e^{i\omega\tau_0}-e^{-i\omega\tau_0}\theta(\tau-\tau_0)\Big]\,.\label{tem}
\end{equation}  
For $\tau >\tau_0$ we reproduce the result (\ref{af2}) and therefore the correct value for the total rate independent of the details of the process of turning on and off the charge. Unlike the rate of scalar radiation (\ref{emra2}) which remains constant after turning on the charge, the electromagnetic rate  ($\sim \partial_\tau |f(\tau,\omega|^2)$) is different from zero only  during  turning on or off the charge where the spatial 1-component of the current is non vanishing. 
For $-\tau_0 < \tau <\tau_0$  and for $\omega \ll a$ the probability for observing a  photon emitted with energy $\omega$ and transverse momentum ${\bf k}_{\perp}$ 
is given by (cf.\,Eq.\,(\ref{chi1}))
\begin{equation}
\big|\chi^1(\tau, \omega, k_{\perp})\big|^2 
= (n_{\omega}+1) \frac{4\pi q_0^2}{\omega a^2\,(2\pi)^{d+1}}
K_1^2\Big(\frac{k_{\perp}}{a}\Big)\,,
\label{chitau}
\end{equation} 
if the initial state contains $n_\omega$ photons. This probability exhibits  a linear divergence with vanishing photon energy  which is further enhanced by the occupation probability of the Unruh heat bath. The divergence  disappears  in the process of switching  off the charge and the finite value of the averaged rate (\ref {hrp})  results. (Details of the time evolution can be studied  using the parametrization (\ref{turn}).)  In the time interval $-\tau_0 < \tau < \tau_0$ a co-accelerated observer measures a  surplus of Rindler photons of small but finite energies  in the Unruh heat bath due to spontaneous emission. The  probability for this surplus diverges like $\omega^{-1}$ with decreasing photon energy.  As far as the origin of this divergence is concerned we cannot rule out at this point that, although independent of the details of the switching on and off processes, it is related to the particular class of parametrizations (\ref{hipa}). Nevertheless, within this class of parametrizations  the divergence is unacceptable. It signals a failure of the perturbative treatment of the coupling to the external current and one might expect on the basis of the non-perturbative expression (\ref{Mmn}) that a summation of higher order terms will cure this problem.  In this context it also may be worthwhile to examine the quantum mechanical consequences of the modification of the retarded Green's function method in the calculation of  the classical radiation field of a uniformly accelerated charge \cite{BOGO55},\cite{BOUL80}.
\subsection{The interaction energy of static charges in accelerated frames}
\begin{figure}
\begin{center}
\includegraphics[width=.4\linewidth]{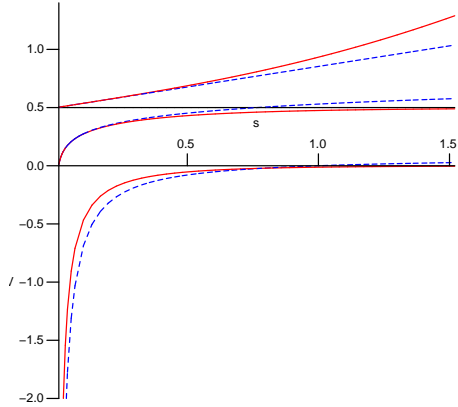}
\caption{The interaction energy (\ref{C13})  (solid lines) for $d=1,2,3$ and the small distance  approximation (\ref{sms}) (dashed lines) as a function of the geodesic distance $s$. Up  to an additive constant the small distance approximation coincides with the interaction energy of static charges in the inertial system. The curves for $d=1$ and $d=2$ are shifted by 0.5}   
\label{pot3}
\end{center}
\end{figure}

In this section we discuss the physics associated with the longitudinal degrees of freedom in Rindler space. As the radiation field, also the Coulomb field in Rindler space generated either by external charges or charged matter fields differs in a characteristic way from that in Minkowski space.   
Starting point for our discussion is the expression (\ref{coul}) for the Coulomb energy of static charge according to which the  interaction energy of  pointlike charges $e_1, e_2, $ located at 
$\xi_1,{\bf x}_{\perp}$ and $\xi_2, {\bf 0}_{\perp}$ respectively, 
is  given by 
\begin{equation}
\label{C12}
V_{C}=-e_1 e_2 
D (\xi_1, {\bf x}_{\perp}, \xi_2, {\bf 0}_{\perp}) \,  ,
\end{equation}
with the Green's function associated with the Laplacian (\ref{lapla}) defined in Eq.\,(\ref{dxxp}). In Rindler space,  after a  Fourier transform in 
${\bf x}_{\perp}$ this Green's function satisfies 
\begin{equation*}
\left ( v\; \frac{\partial}{\partial v} \; \frac{1}{v}\; \frac{\partial}{\partial v} - 
{\bf k}^{2}_{\perp} \right )  g (v, v^{\prime})
= a v \delta (v - v^{\prime}) \, ,\quad v^{(\prime)}=\frac{1}{a} e^{a \xi^{(\prime)}}\, ,
\end{equation*}
with the solution 
\begin{equation}
g (v, v^{\prime})=-a v_{<} I_{1} (k_{\perp} v_{<}) v_{>} K_{1} (k_{\perp} v_{>}) \, .
\end{equation}
 The integration over ${\bf k}_{\perp}$ can be expressed in terms of Legendre functions of the second kind  (cf.\,\cite{RG65}) 
\begin{eqnarray}
D(\xi_1, {\bf x}_{\perp}, \xi_2, {\bf 0}_{\perp}) &=& -a  v_1 v_2 \int \frac{d^{\,d-1}k_{\perp}}{(2 \pi)^{\,d-1}} I_{1} (k_{\perp}v_<) K_{1} (k_{\perp}v_> ) \; e^{i{\bf k}_{\perp}{\bf x}_{\perp}} \nonumber  \\
&=& \frac{i a}{(2 \pi)^{3/2}} \Big[-\pi \zeta^2 \sqrt{u^2-1}\Big]^{-(d-3)/2}  (u^2-1)^{-1/4} Q^{\frac{d}{2}-1}_{\frac{1}{2}} \left(u\right) \, ,
\label{Gelst}
\end{eqnarray}
with 
\begin{equation}
\label{zt}
\zeta^{2} = 2 v_{1} v_{2} \, .
\end{equation} 
The variable $u$ is given by 
\begin{equation}
\label{vu}
u = 1+\frac{s^{2}}{\zeta^2} \, ,
\end{equation}
with 
\begin{equation}
\label{vd}
s^{2} = (v_1 - v_2)^{2} + {\bf x}_{\perp}^{2} \, ,
\end{equation}
the geodesic distance (cf.\,Eq.\,(\ref{sgd})).
With this result, the interaction energy (\ref{C12}) of two oppositely charged sources $e_1=e=-e_2$ can be written as 
\begin{equation}
\label{C13}
a^{-1} V_C = -\frac{e^{- i \frac{\pi}{2} (d-2)}}{2^{3/2} \pi^{d/2}} \; 
\frac{e^{2}}{\zeta^{d-3}} \; (u^{2}-1)^{- (d-2)/4}  \; Q^{\frac{d}{2}-1}_{1/2} (u)\,, \quad d\ge 2.
\end{equation}
The above dimensionless quantity depends on the geodesic distance $s$ (\ref{vd}) and in addition, due to the loss of translational symmetry,  
on the variable  $\zeta $  (\ref{zt}).  If we vary  the geodesic distance, with the ``center of mass'' of the charges 
$$\zeta^2= \frac{2}{a^2} \exp a (\xi_1+\xi_2)$$
kept fixed, the variable $\zeta$ defines the scale in which the dimensionful quantities, the  geodesic distance 
$$\tilde{s} = \frac{s}{\zeta}\,,$$  and the charges $e_1, e_2$  are measured. 
This suggests to introduce the dimensionless coupling constant 
\begin{equation}
\label{rsc}
\tilde{e}^2=\frac{e^{2}}{\zeta^{d-3}}\,,  
\end{equation}
running with the value of the center of mass. 
This definition  is in accordance with the definition (\ref{ruco})  of $\hat{e}$  determined from the scaling properties of the action of the Maxwell field 
$$ \tilde{e}^2 = \Big(\frac{a}{\sqrt{2}}\Big)^{d-3} \hat{e}^2\big(-\frac{1}{2}(\xi_1+\xi_2)\big) \,.$$
The appearance of an effective scale set by one of the coordinates in curved spaces is reminiscent of the  AdS/CFT duality where the  scale is set by the coordinate transverse to the 4-dimensional Minkowski like space \cite{POST01}, \cite{BBSC07}. \\ 
In the limit of small  geodesic distances the above expression reduces to (cf. \cite{EMOT53} Ch. III)
\begin{eqnarray}
a^{-1} V_C   &_{{\longrightarrow \atop \tilde{s} \to 0}}& - \frac{1}{2^{5/2} \pi^{d/2}} \; 
\; \Gamma (\frac{d}{2} - 1)\,\frac{\tilde{e}^2}{\tilde{s}^{d-2}} \, \quad \quad \;d\ge 3\,, \nonumber\\
 &_{{\longrightarrow \atop \tilde{s} \to 0}}& \frac{1}{ 2^{\frac{3}{2}} \pi}\,\tilde{e}^2\,\Big[ \ln \tilde{s} +2 -\frac{5}{2}\ln 2\Big]  \qquad \; \;d=2\,,
\label{sms}
\end{eqnarray}
and we find for large geodesic distances 
\begin{eqnarray}
a^{-1} V{_C}  & _{{\longrightarrow \atop \tilde{s} \to \infty}}& -\frac{\Gamma (\frac{d}{2} + \frac{1}{2})}{8 \pi ^{(d-1)/2}} \; \tilde{e}^{2}  \tilde{s}^{- d-1}\, ,  \qquad \qquad d\ge 2.
\label{las}
\end{eqnarray}
Up to additive constants, the interaction energies coincide for small distances of the charges with the corresponding  expressions for charges at rest in inertial systems while  at large distances the  interaction energies of charges at rest in  accelerated systems are suppressed by three powers of $1/s$. 
For $d=1$, an elementary calculation yields
\begin{eqnarray}
a^{-1} V{_C}  &=& \frac{\tilde{e}^{2}}{4} \, \tilde{s}\sqrt{\tilde{s}^2+2} \,,\qquad  \qquad \qquad \qquad \;\;d=1\,.
\label{vd1}
\end{eqnarray}
Again, the small distance limit coincides with the flat space result.  
 Figure\,\ref{pot3} shows the interaction energies in inertial and accelerated frames  for $d=1,2,3$. 
It  displays the faster  decrease with $s$  of the interaction energies in the  accelerated frame for $d\ge 2$ which in particular  implies a change from the logarithmic rise of the interaction energy     for $d=2$  to a power law decay.  For $d=1$ the increase of the interaction energy is enhanced by the acceleration.\\
We finally remark that the above calculation is independent of the quantum state of the transverse gauge degrees of freedom. We have used Gau\ss\ \!\!\!\! ' law which is valid for any physical state. In particular our calculation applies equally to the ground state in the inertial and to the ground state in the accelerated system.

It is instructive to compare the interaction energy of static charges coupled to the Maxwell field  with that of static sources coupled to a massless scalar field. According to Eqs.\,(\ref{hscri}) and (\ref{EQMin}), the energy arising from the coupling of static sources to the scalar field is given by 
$$ h^{sc} = \frac{1}{2} \int e^{2a\xi}\ d\xi d^{d-1}x_{\perp}\int e^{2a\xi^{\prime}}\ d\xi^{\prime} d^{d-1}x^{\prime}_{\perp}\rho(\xi,{\bf x}_{\perp}) D_s(\xi,{\bf x}_{\perp},\xi^{\prime},{\bf x}^{\prime}_{\perp}) \rho(\xi^{\prime},{\bf x}^{\prime}_{\perp})\,, $$
with the scalar Green's function (cf.\,Eq.\,(\ref{scala}))
 $$D_s(\xi,{\bf x}_{\perp},\xi^{\prime},{\bf x}^{\prime}_{\perp})= \langle \xi,{\bf x}_{\perp}| \frac{1}{\Delta_s} |\xi^{\prime},{\bf x}^{\prime}_{\perp}\rangle\,.$$
Therefore the interaction energy of two pointlike sources of strength $q_1, q_2$
reads
\begin{equation}
\label{inerg}
V_{sc}=q_1 q_2 
D_s (\xi_1, {\bf x}_{\perp}, \xi_2, {\bf 0}_{\perp}) \,  .
\end{equation} 
Unlike in Minkowski space, interaction energies of static sources coupled to scalar or vector fields are not given in terms of  the same Green's functions. They exhibit different spatial dependences.  
The  Laplacians $\Delta_s$ (\ref{scala}) for scalar and $\Delta$ (\ref{lapla}) for vector fields are related to each other by  
$$\Delta_s = \Delta +\frac{2}{v}\, \partial_{v}$$
and a similar  calculation as in (\ref{Gelst})  yields the following ratio of propagators
\begin{equation}
\frac{D_s (\xi_1, {\bf x}_{\perp}, \xi_2, {\bf 0}_{\perp}) }{D (\xi_1, {\bf x}_{\perp}, \xi_2, {\bf 0}_{\perp}) } = 
\frac{1}{a^{2} v_{1} v_{2}} \; \frac{Q^{d/2-1}_{-1/2} (u)}{Q^{d/2-1}_{1/2} (u)}= e^{-a(\xi_1+\xi_2)} (1+ \tilde{s}^2+\tilde{s}\sqrt{2+\tilde{s}^2})\,.
\label{vatv}
\end{equation}
We observe that  the dependence of the interaction energy of scalar sources on the distance ($\tilde{s}$) coincides at small distances  with  that of static charges, and thus with the flat space result. At large distances the  interaction energy is suppressed  with respect to  the flat space result  by only one power of $\tilde{s}$.

For $d=3$, as for all odd spatial dimensions, the interaction  energy can be expressed in terms of elementary functions. We find from (\ref{C13})  (cf.\cite{RG65})
\begin{equation}
\label{vcoul}
V_C= \frac{e^2}{4\pi}  a \left ( 1 - \frac{1+\tilde{s}^2}{\sqrt{2\tilde{s}^{2} +\tilde{s}^4}}  \right )\, .
\end{equation}
The modification of the asymptotics by acceleration would change significantly the spectrum of atoms  or molecules. In particular one would expect that a hydrogen atom should exhibit only a finite number of bound states separated from the threshold by a finite gap. The effect of the acceleration should be most strongly pronounced in  ``Rydberg'' atoms i.e. atoms in highly excited and very weakly bound states, (cf.\cite{PINT93}   for a study of Rydberg atoms in the Schwarzschild metric). In  low lying states and in the perturbative limit $a a_B \ll 1$,  with  $a_B$ the Bohr radius, the modifications of the kinetic energy  are actually dominant. This can be easily verified using the following  Hamiltonian for the hydrogen atom in Rindler space
 \begin{equation}
 H=\frac{1}{2} m (e^{2a\xi}-1) 
 - \frac{1}{2m} (\partial^{2}_{\xi} + e^{2 a \xi}\partial^{2}_{\perp})  +e^2 D(\xi,{\bf x}_{\perp}, 0,{\bf 0}_{\perp})\, .
\label{SCHEQMA}
\end{equation} 
For energies small in comparison to the electron's rest mass, this effective Hamiltonian is derived  from the scalar field action (\ref{qedsc}) and  the expression (\ref{fesc2}) for the interaction energy of a static proton with the charge density of the (scalar) electron (\ref{j0sc}).   Besides ${\bf x}_{\perp}={\bf 0}$, for simplicity $\xi_{\text{proton}}=0$ has been assumed.

The suppression of long range forces by acceleration may also be of relevance for Yang-Mills theories and QCD and could be taken as an indication for deconfinement by acceleration.  It has been actually argued \cite{KATU05} that the Unruh effect provides a new mechanism for thermalization through the deceleration  occurring when two relativistic heavy ions collide and, for instance, could  trigger restoration of chiral symmetry. On the basis of our results one might conjecture that in Rindler space a confined phase in Yang-Mills theories or a chirally broken phase in QCD does  not exist at all. The thermal distribution of photons (cf. Eq.\,(\ref{urtmax})) or of massive scalars (cf.\,Eq.\,(\ref{urt2})) is independent of the value of the transverse momenta and mass of the ``Rindler particles'' resulting in an ultraviolet divergence of the entropy which requires regularization \cite{SULE05}.  This  degeneracy can be expected to remain true  as far as the elementary excitations of  interacting theories  (cf.\,\cite{UNRU84}, \cite{SULE05}), for instance the ``Rindler pions'' in QCD, or ``Rindler glueballs'' in Yang-Mills theories  are concerned. In turn, the existence of a confined phase in Yang-Mills theories in Rindler spaces appears to be incompatible with the expected degeneracy of the ground state, the large contributions of the  one-particle states to the entropy or the missing gap in the spectrum. As far as  the issue of thermalization in relativistic heavy ion collisions is concerned, it is not straightforward to apply  these considerations  without having studied  the changes in the dynamics if the acceleration does not last forever.  
\section{Conclusions}
In this work we have presented the canonical quantization of gauge fields in static space-times and have applied the formalism to the electromagnetic field in Rindler spaces. The existence of a timelike Killing vector in static space-times makes the quantization of gauge fields in the Weyl gauge very similar to that in flat space-time. For electromagnetic fields,  we have developed a general scheme for treatment of the Gau\ss\ law which appears in the Weyl gauge as a constraint on the physical states. The two crucial elements in this construction are the definition of metric dependent transverse and longitudinal fields and the decoupling of the longitudinal degrees of freedom. This decoupling is achieved by a unitary  gauge fixing transformation and enables explicit determination of the stationary states of the longitudinal degrees of freedom. We have shown that in general static space-time the time component of currents either external or generated by charged matter fields is coupled to the radiation field and have discussed the possibility   to eliminate this coupling by a suitably chosen coordinate transformation. A complete construction of the stationary states and an explicit non-perturbative solution of initial value problems has been given in case the electromagnetic field is coupled to external currents only. As far as further formal developments are concerned it would be interesting to consider alternatives to the elimination of longitudinal degrees of freedom.  The decoupling of these degrees of freedom is one particular option for resolving the Gau\ss\ law constraint. As in flat space and depending on the metric or on prescribed boundary conditions elimination of other degrees of freedom may be more practical, for instance, the elimination of a particular component of the gauge field (cf.\,\cite{LNOT94}). Such an alternative could be promising for curved space-times like Rindler or AdS spaces where one of the spatial directions is singled out and, most likely,    will be necessary  for resolving the Gau\ss\ law in non-Abelian gauge theories. Furthermore it appears promising to extend our formal development and include stationary space-times \cite{FULL89}\cite{WALD94} where the basis of the canonical quantization in the Weyl gauge, the existence of a time-like Killing vector, persists. If successful, such an extension would permit applications to metrics like the Kerr metric.

Using the tools developed for gauge fields in general static space-times we have studied  the dynamics of the electromagnetic field in Rindler spaces. We have constructed the normal modes of the transverse gauge fields and expressed the Hamiltonian in terms of the associated creation and annihilation operators. We have  identified the symmetry which is responsible for the degeneracy of the energy eigenvalues with respect to the momentum components perpendicular to the direction of the homogeneous gravitational field and have shown that the Rindler space Hamiltonians of a non-interacting scalar and of the Maxwell field are  invariant under scale transformations. We have extended these symmetry considerations to scalar and spinor QED with massless charged matter fields and to SU(N) QCD with massless quarks and found that in 4-dimensional Rindler space these interacting systems exhibit at the tree level this exact symmetry. It remains to be seen which signatures of this symmetry will be left after accounting for the quantum effects of the interactions. 

Observed in a  uniformly accelerated frame, Minkowski space-time appears as a Rindler space. We have studied in detail the ensuing relation of quantized electromagnetic fields in Rindler and Minkowski spaces and have established the relation between the creation and annihilation operators of the corresponding transverse fields.  We have derived non-perturbative expressions for transition rates for photon emission and absorption induced by external currents and, for the sake of comparison, the corresponding transition rates for massless scalar fields. The absence of radiation of a uniformly  accelerated charge  in the co-accelerated  frame has been an important issue in the discussion of the Unruh effect. It has been argued that the rate of response, i.e. the sum of absorption and emission rate, must be the same in inertial and accelerated frames. This conjecture is supported by our results for both scalar and electromagnetic sources. The time evolution of emission and absorption amplitudes differ significantly for scalar and electromagnetic radiation. In particular, after turning on the  charge, the  amplitude for photon emission exhibits an infrared divergence which disappears after the charge is turned off while the rate of scalar radiation remains constant after turning on the source. Additional investigations, in particular of contributions from higher order perturbation theory are necessary.  For further clarification of bremsstrahlung processes, an analysis of the dynamics of appropriately constructed Unruh-DeWitt detectors for gauge fields seems to be mandatory. So far almost all investigations of the response of uniformly accelerated detectors have dealt with scalar radiation (for a survey cf.\,\cite{CRHM07}). A dipole-detector coupled to the electromagnetic field has been introduced in \cite{TAGA86} and used in a study of an accelerated hydrogen atom \cite{PASS98} for a comparison of atomic level shifts arising from coupling to  scalar and gauge fields respectively.  A similar Hamiltonian, for instance of a harmonic oscillator with a dipole coupling to the electric field,  could be used as a model of an ``electromagnetic'' detector. For such a detector coupled to a scalar field non-perturbative results are available \cite{LIHU06}. Our method of decoupling  gauge fields and external currents  by unitary transformations is actually very similar to the method of solution applied in \cite{LIHU06} and should be readily generalizable as long as the dynamics  remains  bilinear in all  degrees of freedom.

Besides the initial value problem associated with bremsstrahlung, Rindler spaces offer the opportunity for analytical studies of other non-trivial properties of quantized gauge fields. We have computed the photon condensate which appears  in a uniformly accelerated system in space-time dimensions different from four and have investigated the electrostatic interaction of two charges at rest in Rindler space. While in comparison with  Minkowski space the  short distance behavior of the  interaction energy is the same, its  asymptotics is suppressed by three powers of the distance. For the interaction energy of two sources coupled to a massless scalar field we obtain in comparison with Minkowski space again the same  small distance behavior and  also an asymptotic  suppression however by one power of the distance only. As for the density of states  of the transverse degrees of freedom we thus also find  a significant difference between scalar and longitudinal degrees of freedom. The asymptotic suppression of the electrostatic energy changes qualitatively the spectrum of the hydrogen atom. Together with the expected absence of a gap in the spectrum and the degeneracy of the one-particle states this asymptotic suppression of the electrostatic energy can be taken as a hint that  non-Abelian gauge theories in Rindler spaces may not exhibit confinement. 

Although we have concentrated the discussion on gauge fields in Rindler spaces or equivalently gauge fields in uniformly accelerated systems, extensions to general stationary accelerated frames  (\cite{LEPF81,KOLE04}) appear to be feasible and of particular interest for the experimental verification of the Unruh effect. One also can expect  analytical results from a study of the Unruh effect for gauge fields in AdS spaces (cf.\,\cite{DELE97}).   
\begin{acknowledgments} 
It is a pleasure to express our sincere thanks to  Tetsuo Matsui for his collaboration in the early phase of this project. F.\,L. is grateful for support by a fellowship of the Japan Society for the Promotion of Science and for the hospitality at the Institute of Physics  at the University of Tokyo, Komaba. K.\,O. and K.\,Y. would like to thank their collaborator F.\,L. for the support 
of stay and the hospitality at the University of Erlangen. 
K.Y. is also supported by the Grant-in-Aid for the Scientific Research of
JSPS No.15540288 and No.19540306 and thanks Wolfgang Bentz who is the 
principal investigator of the latter. 
\end{acknowledgments} 

\end{document}